\documentclass[12pt]{article}
\usepackage{geometry}
\usepackage{authblk} 
\usepackage{lineno} 
\usepackage{microtype}
\usepackage{amsbsy} 
\usepackage{amssymb}
\usepackage{amsmath}
\usepackage[round]{natbib}
\usepackage{csvsimple}
\usepackage{footnote}
\usepackage{tablefootnote}
\usepackage{array}
\usepackage{booktabs}
\usepackage{adjustbox}
\usepackage{multirow}
\usepackage{arydshln}
\usepackage{pdfpages}
\usepackage{graphicx}
\usepackage{caption}
\usepackage[justification=justified,singlelinecheck=false]{subcaption}
\usepackage{draftwatermark}
\usepackage{setspace}

\SetWatermarkText{Submitted to Bulletin of\\Earthquake Engineering}
\SetWatermarkScale{3}

\title{A Non-ergodic Effective Amplitude Ground-Motion Model for California}

\author[1]{Grigorios Lavrentiadis\thanks{glavrent@berkeley.edu}}
\author[1]{Norman A. Abrahamson\thanks{abrahamson@berkeley.edu }}
\author[2]{Nicolas M. Kuehn\thanks{kuehn@g.ucla.edu }}
\affil[1]{Department of Civil Engineering, University of California, Berkeley}
\affil[2]{B. John Garrick Institute for the Risk Sciences, University of California, Los Angeles}

\begin{document}

\maketitle

\begin{abstract}
A new non-ergodic ground-motion model (GMM) for effective amplitude spectral ($EAS$) values for California is presented in this study. 
$EAS$, which is defined in \cite{Goulet2018}, is a smoothed rotation-independent Fourier amplitude spectrum of the two horizontal components of an acceleration time history. 
The main motivation for developing a non-ergodic $EAS$ GMM, rather than a spectral acceleration GMM, is that the scaling of $EAS$ does not depend on spectral shape, and therefore, the more frequent small magnitude events can be used in the estimation of the non-ergodic terms. 

The model is developed using the California subset of the NGAWest2 dataset \cite{Ancheta2014a}.
The \cite{Bayless2019a} (BA18) ergodic $EAS$ GMM was used as backbone to constrain the average source, path, and site scaling. 
The non-ergodic GMM is formulated as a Bayesian hierarchical model: the non-ergodic source and site terms are modeled as spatially varying coefficients following the approach of \cite{Landwehr2016}, and the non-ergodic path effects are captured by the cell-specific anelastic attenuation attenuation following the approach of \cite{Dawood2013}.
Close to stations and past events, the mean values of the non-ergodic terms deviate from zero to capture the systematic effects and their epistemic uncertainty is small. 
In areas with sparse data, the epistemic uncertainty of the non-ergodic terms is large, as the systematic effects cannot be determined.  

The non-ergodic total aleatory standard deviation is approximately $30$ to $40\%$ smaller than the total aleatory standard deviation of BA18. 
This reduction in the aleatory variability has a significant impact on hazard calculations at large return periods. 
The epistemic uncertainty of the ground motion predictions is small in areas close to stations and past events.

\end{abstract}

\section{Introduction} \label{sec:intro}
Probabilistic seismic hazard analysis (PSHA) estimates the annual rate of exceeding a ground-motion parameter at a site of interest.
It typically breaks the problem in two parts: the seismic source characterization and the ground-motion model (GMM). The first part defines the rate, the magnitude distribution, and the location distribution of earthquakes in a region, and the second part, which is the focus of this study, provides the probability of exceeding the ground-motion for a specific earthquake scenario. 
Most GMM are derived empirically using strong-motion datasets and define the distribution of the ground-motion parameter as a function of source, path, and site parameters such as magnitude ($M$), closest rupture distance ($R_{rup}$) and time-averaged shear-wave velocity in the top $30m$ ($V_{S30}$).
Traditionally, due to the scarcity of regional data, GMM were developed under the ergodic approach which assumes the statistical properties of the ground-motion parameter do not change in space \citep{Anderson1999}. 
These ergodic models tend to have large aleatory variability as they treat some of the systematic effects for a specific site/source location as random variability that can occur anywhere. 
Examples of models that have been developed under this approach are the NGA-West GMMs for California \citep{Abrahamson2008}, and the European GMMs derived from the RESORCE database \citep{Douglas2014}; however, as more data are collected, the ergodic assumption can be relaxed, and repeatable effects related to the source, path and site can be properly modeled, which leads to a decrease in the aleatory variability.
This reduction has a large impact on the hazard at large return periods because, the ground-motion aleatory variability controls the slope of the hazard curves which has a large influence on the hazard at large return periods. 
\cite{AlAtik2010} describes how the aleatory variability of an ergodic model can be separated into epistemic uncertainty related to the systematic source, path, and site terms, and the reduced aleatory variability of a non-ergodic GMM.
The epistemic uncertainty refers to the range by which the non-ergodic terms vary in areas with no available data to constrain them. 
In areas with data from past earthquakes, the non-ergodic terms can be estimated, and their epistemic uncertainty can be reduced.

The first step in this new paradigm was to create a regional GMM or a global GMM with some regionalized terms.
Regional GMMs are developed with smaller regional data-sets, for instance \cite{Akkar2010} for Turkey, \cite{Bindi2011} for Italy, \cite{Bragato2005} for Eastern Alps, and \cite{Danciu2007} for Greece.
These models have smaller aleatory variability than global ergodic GMMs, but they suffer from weaker constrains on the scaling due to the smaller size of the regressed data sets.
Global GMMs with regionalized terms are developed with large global datasets, the same way ergodic GMM are developed but with the difference that some of the scaling terms are estimated separately for each region.
The NGAWest2 GMM \citep{Bozorgnia2014} followed this approach.
For example, in \cite{Abrahamson2014}, both the $V_{S30}$ and anelastic attenuation scaling terms were regionalized: they have a different set of coefficients for Califonia, China, Japan, and Taiwan. 

Partially non-ergodic GMMs that only capture the systematic site effects, known as single station GMMs, lead to an approximately $30\%$ reduction in the aleatory variance compared to an ergodic GMM \citep{Coppersmith2014, Bommer2015, Tromans2019}.
Similarly, other researchers have developed partially non-ergodic GMMs that capture repeatable effects related to the source \citep{Yagoda-Biran2015}, path \citep{Dawood2013, Kuehn2019}, and single source/single site \citep{Hiemer2011}.
Fully non-ergodic GMMs include non-ergodic terms to capture simultaneously all the aforementioned systematic effects (source, path, and site); these type of models have $60$ to $70\%$ smaller aleatory variance than ergodic GMMs \citep{Lin2011, Landwehr2016, Abrahamson2019}.

The model presented in this study is a fully non-ergodic GMM that captures the systematic effects of the source, site, and anelastic attenuation from the path.
It is developed as spatially varying coefficient model (VCM), following the methodology used in \cite{Bussas2017} and \cite{Landwehr2016}.
The non-ergodic anelastic attenuation is modeled with the cell-specific anelastic attenuation similar to \cite{Dawood2013} and \cite{Abrahamson2019}.
VCMs impose a spatial correlation on the coefficients, so that they vary continuously from location to location; alternative methods to impose the spatial correlation are: splines, and neural networks.
In the cell-specific anelastic attenuation approach, the domain of interest is broken into cells which all have their own attenuation coefficient.
The attenuation along a path from a source to site is equal to the sum of attenuation of the cells that it traverses. 

One distinction of the current model from other non-ergodic GMMs is that it is developed for effective amplitude spectral ($EAS$) values instead of response spectral accelerations ($PSA$).
This is done because it is easier for the $EAS$ non-ergodic effects estimated from small events to be transferred to large magnitude earthquakes than for the $PSA$ non-ergodic effects.
The response of a small-period single-degree-of-freedom oscillator (SDOF) is affected by the entire frequency content of the ground motion, as amplitude of the short-period ground motion is too small control the SDOF response.  
This makes the short-period $PSA$ sensitive to the ground motion at frequencies near the peak in the $FAS$ due to their higher amplitudes. 
The frequency content and period of the $FAS$ peak change with magnitude: as the magnitude of an earthquake increases, the period of the peak also increases, making the scaling of the small-period $PSA$ magnitude dependent.
It is this magnitude dependence of the $PSA$ scaling that makes the short-period non-ergodic terms of the small magnitudes not transferable to large magnitude events without modification.
A detailed discussion of the scaling of $FAS$ and $PSA$ is given by \cite{Bora2016}.


As an example, the response spectrum of an $M 3$ event has its peak at $T=0.1 sec$, whereas the $M 7$ response spectrum from has its peak at $T=0.3 sec$; a sketch illustrating their different spectral shapes is provided in the electronic supplement.
Due to this spectral shape difference, the $PGA$ scaling will be consistent with the scaling of the $PSA(T=0.1 sec)$ at small magnitudes and with the scaling of the $PSA(T=0.3 sec)$ at large magnitudes.
For example, to properly capture the magnitude dependence in a $PSA$ GMM, the $V_{S30}$ coefficient for $PGA$ would be close to the $V_{S30}$ coefficient for $T=0.1 sec$ at small magnitudes, and it should gradually shift towards the $V_{S30}$ coefficient for $T=0.3 sec$ as magnitude increases. 
This is consistent with \cite{Stafford2017}, who showed that the linear site amplification factors are magnitude and distance dependent.

The EAS is defined in \cite{Goulet2018} as the smoothed power-averaged Fourier amplitude spectrum ($FAS$) of the two horizontal components.
The EAS does not suffer from the same problem as $PSA$ because the Fourier transform is a linear operation, and the scaling coefficients and non-ergodic terms estimated from small magnitude earthquakes can be applied directly to large events for linear source, path, and site effects.  
To ensure that the proposed model extrapolates reasonably to scenarios outside the range of events in the regression data set, we use the \cite{Bayless2019a} ergodic $EA$S GMM (BA18) as a backbone model; 
we selected BA18 because it was developed on a large global data-set, and it includes seismological constraints to avoid oversaturation at short distances and large magnitudes. 
The non-ergodic terms and cell-specific anelastic attenuation coefficients were estimated with the total residuals of BA18 from the NGAWest2 California subset.


Lastly, one common comment regarding the usage of an EAS GMM is that, in most seismic design methods, the intensity of the ground-motion shaking is specified in terms of $PSA$ and not $EAS$.
We can use an $EAS$ GMM and Random Vibration Theory (RVT) to compute the equivalent $PSA$.
RVT uses FAS and the duration of an SDOF oscillator response to a ground motion to compute the root-mean-square amplitude of the response ($x_{rms}$), based on Parseval's theorem.
The spectral acceleration, which is the peak response of the SDOF, is estimated by the product of $x_{rms}$ and the peak factor. 
\cite{Boore1983} used RVT with FAS from seismological theory to calculate the equivalent $PSA$, and \cite{Bora2015, Bora2019} derived a duration model which, when used with a \cite{Brune1970, Brune1971} omega-squared FAS model, gives predictions that are consistent with the NGAWest2 data set.
Converting the non-ergodic $EAS$ GMM into an equivalent $PSA$ GMM is not in the scope of this paper; it is covered in the second part of this study. 
In this approach, the median estimate of the non-ergodic $PSA$ GMM is based on the median estimate of the $EAS$ GMM and RVT, while the non-ergodic aleatory variability is estimated empirically using the ground-motion observations. 

\section{Ground-Motion Data} \label{sec:data}
\begin{figure} 
    \centering
    \begin{subfigure}[t]{0.45\textwidth}
        \includegraphics[width = \textwidth]{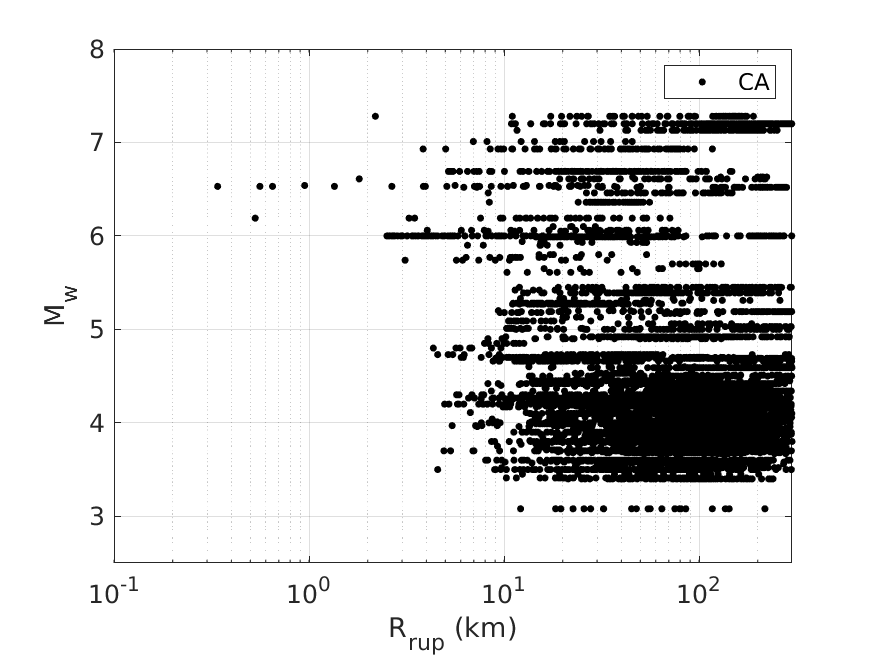}
    \end{subfigure}  
    \begin{subfigure}[t]{0.45\textwidth}
        \includegraphics[width = \textwidth]{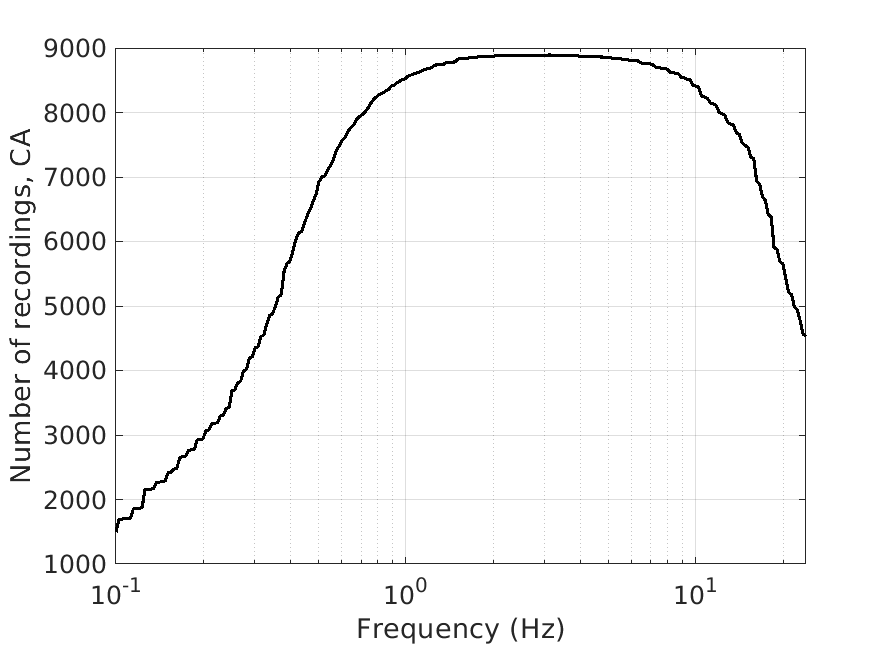}
    \end{subfigure}
    \caption{Selected data from the NGAWest2 database. The left figure shows Magnitude - Distance distribution of the subset used in the regression analysis. The right figure the number of recordings per frequency used in the regression analysis.}
    \label{fig:dataset_nga2CA}
\end{figure}

The NGAWest2 data-base \citep{Ancheta2014} includes more than 21000 recordings covering a magnitude range from $3$ to $7.9$ and a closest distance range ($R_{rup}$) from $0.05$ to $1500$ km.
For this study, a subset of this data-base was used which included the earthquakes and stations located in California, western Nevada, and northern Mexico.
The recordings that were identified as questionable in \cite{Abrahamson2014} were not used in the regressions.
The final data-set contains $8916$ recordings from $188$ earthquakes recorded at $1497$ stations.
Figure \ref{fig:dataset_nga2CA} shows the magnitude-distance distribution of the data and the number of recordings per frequency.
The earthquake magnitudes in the selected data range from $3.1$ to $7.3$ and the distances range from $0.1$ to $300$ km. 
The usable frequency range of the majority of the recordings spans from $1$ and $10 Hz$.

\begin{figure}
    \centering
    \includegraphics[width = 0.5 \textwidth]{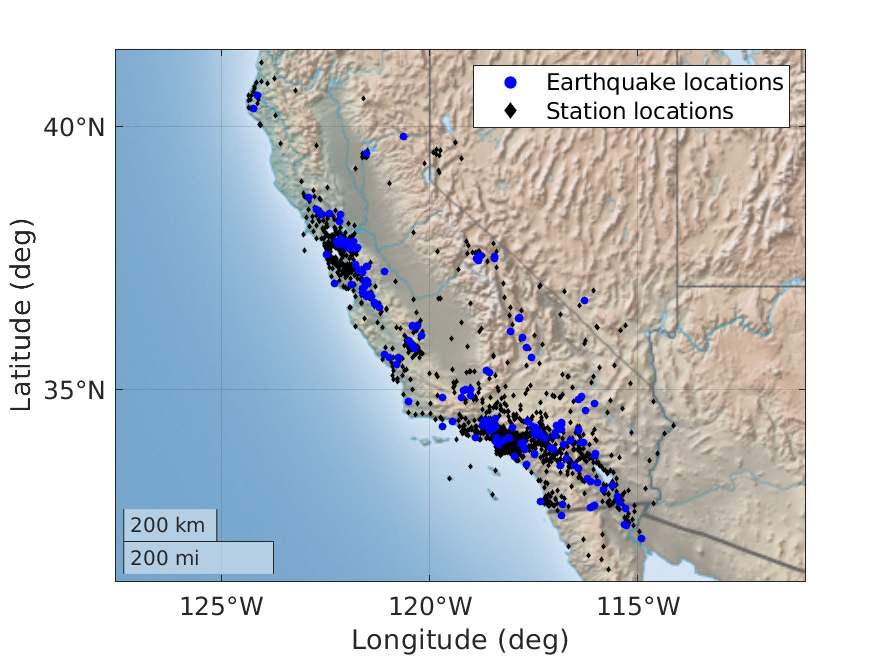}
    \caption{Spatial distribution for earthquakes and stations used in this study.  }
    \label{fig:dataset_spat_dist}
\end{figure}

Figure \ref{fig:dataset_spat_dist} shows the spatial distribution of the data: most of the stations are located in Los Angeles, Bay Area, and San Diego metropolitan areas, whereas in less populated areas, such as northern-eastern California the spatial density of the stations is lower.
This difference in the density of stations has a large impact on the distribution of epistemic uncertainty of the systematic site effects: the epistemic uncertainty is higher is areas with lower station density making a case for expanding the strong-motion networks in these regions near critical infrastructure.
In this study, the location of the earthquakes and stations is defined in kilometers in UTM coordinates; the longitude/latitude coordinates were transformed to UTM coordinates using the WGS84 reference ellipsoid and 11S UTM zone.

\section{Non-ergodic Model Development} \label{sec:gmm_dev}
Rather than developing the non-ergodic GMM from scratch, the \cite{Bayless2019a} (BA18) ergodic EAS GMM was used as a backbone model to describe the average ground-motion scaling. 
The main reasons for that decision were that: i) the local data may non be adequate to estimate the scaling of complex terms, and ii) the adoption of the constraints built into BA18 ensures that it extrapolates properly outside the range of data.

\subsection{Functional Form} \label{sec:gmm_fun}

Typically, a GMM (Equation \eqref{eq:eas_fun}) is composed of the median model ($y_{med}$) and the aleatory variability.
The median model describes the average scaling of a ground-motion parameter with magnitude, distance, site conditions, etc.
The aleatory variability describes the misfit between a ground-motion observation and $y_{med}$ which is related to the true or apparent (due to simplified modeling) stochastic behavior of the source, site, and path. 
It is typically expressed as the sum of the between-event ($\delta B_e$) and within-event ($\delta W_{e,s}$) terms. 
$\delta B_{e,s}$ describes average shift of the ground motion for an earthquake, $e$, from $y_{med}$, and $\delta W_{e,s}$ describes the variability of the ground motion at site, $s$ from the median ground motion of that earthquake.

\begin{equation} \label{eq:eas_fun}
    ln(EAS) = y_{med} + \delta B_e + \delta W_{e,s} 
\end{equation}

The median EAS model of \cite{Bayless2019a} is formulated as:
\begin{equation}
    y_{med} = f_M + f_R + f_S + f_{ztor} + f_{NM} + f_{Z1}
\end{equation}
\noindent where $f_M$ is the magnitude scaling, $f_R$ is the path scaling, $f_S$ is the site scaling, $f_{ztor}$ is the top of rupture scaling, $f_{NM}$ is the normal-fault scaling, and $f_{Z1}$ is the basin thickness scaling. 
The $f_M$, $f_R$ and $f_S$ terms were modified to include the additional non-ergodic terms, whereas $f_{ztor}$, $f_{NM}$ and $f_{Z1}$ were kept fixed.
The different terms and coefficients are described in detail in \cite{Bayless2019a} and \cite{ChiouYoungs2014}. 

\subsubsection{Magnitude Scaling} \label{sec:gmm_mag}

The $f_M$ in the non-ergodic model is:

\begin{equation}
    f_M = c_{1} + c_2 (M-6) + \frac{c_2-c_3}{c_n} \ln(1+ e^{c_n (c_M - M)}) + \delta c_0 + \delta c_{0,e} + \delta c_{1,e} (\vec{x}_{e})
\end{equation}

\noindent in which $c_{1}$ is the intercept of the model, $c_2$ controls the magnitude scaling for large magnitudes where the corner frequency is smaller than the frequency of interest, $(c_2-c_3) / c_n$ describes the magnitude scaling at small magnitudes, where the corner-frequency is larger than the frequency of interest, $c_n$ controls the width of the transition zone between small and large magnitude earthquakes, and $c_M$ is the magnitude at the center of the transition zone.
A sketch of the magnitude scaling is provided in the electronic supplement.

The additional coefficients in the non-ergodic model are $\delta c_0$, $\delta c_{0,e}$, and $\delta c_{1,e}$.
$\delta c_0$ is added to allow a small constant shift in the non-ergodic model due the difference in the weighting of residuals between the ergodic and non-ergodic GMMs.
$\delta c_{1,e}$ is a function of the earthquake coordinates, $\vec{x}_{e}$, and is intended to capture repeatable non-ergodic effects related to the source location. 
For instance, regions with a higher than average median stress drop will have higher than average median ground-motions resulting in a positive $\delta c_{1,e}$. 
$\delta c_{0,e}$ is a regional term that corrects a potential bias in the magnitude estimation of small earthquakes between northern and southern California.
$\delta c_{0,e}$ is applied to earthquakes less than magnitude $5$ and frequencies less than $5 Hz$.
The vertices which define the polygons for the northern and southern CA subregions are summarized in Table \ref{tab:delc0NS}, the border between northern and southern CA corresponds approximately to the boundary between the Northern California Seismic Network (NCSN) and Southern California Seismic Network (SCSN).


The $\delta c_{0,e}$ term addresses potential inconsistencies in the reported magnitudes of small earthquakes between northern and southern CA because the NCSN/SCSN boundary was also used in the NGAWest2 dataset \cite{Ancheta2014a} for the selecting the catalog to use for the source parameters (magnitude, hypocenter location, etc.) for small-to-moderate (less than $M~5$) earthquakes.
If a small earthquake was located north of the boundary, the NCSN catalog was used for the source parameters, whereas if a small earthquake was located south of the boundary, the SCSN/CIT catalog was used for the source parameters. 
Preliminary regressions that did not include the $\delta c_{0,e}$ term showed significant differences in $\delta c_{1,e}$ between northern and southern CA at small frequencies. 
It was found that these differences were caused by a noticeable bias in the total residuals of BA18 between northern and southern CA for small magnitude events. 
\cite{Chiou2010} made a similar observation for the total residuals of CY08: they found a regional difference in median ground-motion amplitude between north and south CA earthquakes which was more noticeable at small magnitude events.
The results in section \ref{sec:results} show that the difference in the median ground-motion of small events between northern and southern CA is approximately $0.4$ in natural-log units at frequencies between $0.2$ and $5.0~Hz$.
At frequencies well below the corner frequency, a unit change in magnitude leads to a factor of $32$ change in the amplitude of the ground motion; thus, a $0.4$ natural-log difference in ground motion can be caused by a $0.11$ bias in the magnitude estimation between the NCSN and SCSN networks, which could be due to different assumptions in the velocity models or other input parameters used to determine the magnitude of an event.
Given the assumption of magnitude bias, the $\delta c_{0,e}$ term is not applied to events less than $M~5$.
The source parameters of the moderate-to-large events in NGAWest2 were estimated with fault inversions using data from global networks, and therefore, the potential magnitude bias between the NCSN and SCSN networks is not applicable to events larger than $M~5$. 
However, this issue should be further investigated in future studies to find the exact cause of this regional difference. 

\begin{table}
	\caption {Vertices of Northern and Southern CA regions for $\delta c_{0,e}$}
	\centering
	\label{tab:delc0NS}
	\begin{tabular}{c c  c c}
	    \hline\noalign{\smallskip}
	    \multicolumn{2}{c}{Northern CA}   & \multicolumn{2}{c}{Southern CA} \\
	    lat. (deg)  & lon. (deg)        & lat. (deg)  & lon. (deg) \\
	    \noalign{\smallskip}\hline\noalign{\smallskip}
	    34.5175     & -121.5250         & 37.9775     & -116.6225 \\
	    39.8384     & -125.2341         & 35.2944     & -113.4142 \\
	    41.3595     & -124.1684         & 31.4772     & -115.0250 \\
	    41.3995     & -120.7227         & 31.0082     & -117.6898 \\
	    37.9775     & -116.6225         & 34.5175     & -121.5250 \\
	    \noalign{\smallskip}\hline\noalign{\smallskip}
	 \end{tabular}
\end{table}
	
\subsubsection{Path Scaling} \label{sec:gmm_path}

The functional form for $f_P$ is:
\begin{equation}
\begin{aligned}
    f_P =& c_4 \ln \left( R_{rup} + c_5 \cosh \left( c_6 \max( M - c_{hm},0) \right) \right) + (-0.5 - c_4) \ln(\hat{R}) \\
         &+ \vec{c}_{ca,p}( \vec{x}_{c}) \cdot \Delta \vec{R}
\end{aligned}
\end{equation}
\noindent where $\hat{R} = \sqrt{R_{rup}^2 + 50^2}$. 
The coefficient $c_4$, which corresponds to the geometrical attenuation, manages how the ground motion attenuates at short distances. 
The coefficient $c_5$ describes the short-distance saturation, this term increases the effective  distance for large magnitudes to capture the finite-fault effects (i.e. as the earthquake magnitude increases, the size of the rupture increases resulting in more seismic waves coming from more distant segments of the rupture leading to a larger effective rupture distance).
Coefficients $c_4$ and $c_6$ control the magnitude saturation at short distances. 
Full saturation at zero distance is achieved when $c2 = - c_4 c_6$ is satisfied, non full saturation (i.e. positive magnitude scaling) is achieved when $c_2 > - c_4 c_6$.
At distances greater than $50 km$, the term $(-0.5 - c_4$) cancels the empirically estimated geometrical attenuation and fixes it to $0.5$ which is the theoretical geometrical attenuation of surface waves.
To maintain proper distance scaling and magnitude saturation in the non-ergodic model, all the aforementioned coefficients were fixed to their $BA18$ values.
The non-ergodic distance scaling is captured with the cell-specific anelastic attenuation.

The anelastic attenuation is modeled with the cell-specific anelastic approach, first proposed by \cite{Dawood2013} and then extended by \cite{Kuehn2019} and \cite{Abrahamson2019}. In this method, the states of California and Nevada are broken into $25 \times 25 km$ cells and, for each record, the ray path which connects the site ($\vec{x}_{s}$) to the closest point on the rupture ($\vec{x}_{cls}$) is broken into cell-path segments ($\Delta R_i$) which are lengths of the ray within each cell. 
For each record, the sum of cell-path segment lengths $\sum_{i=0}^{N_c} \Delta R_i$, is equal to $R_{rup}$.

\begin{figure} 
    \centering
    \begin{subfigure}[t]{0.45\textwidth}
        \raisebox{-\height}{ 
            \includegraphics[width = .78\textwidth]{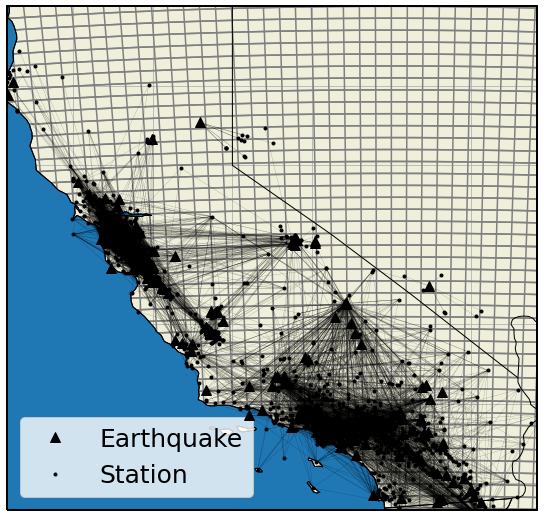} }
    \end{subfigure}  
    \begin{subfigure}[t]{0.45\textwidth}
        \raisebox{-\height}{ 
            \includegraphics[width = \textwidth]{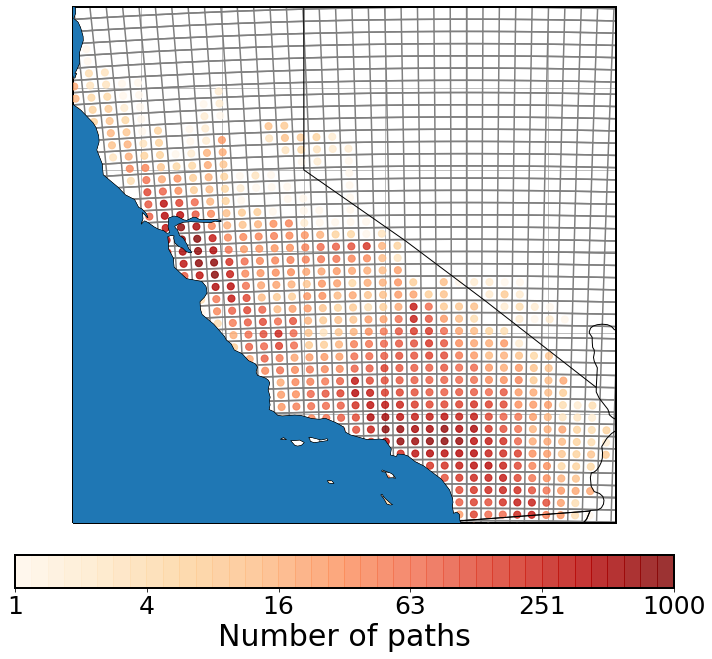} }
    \end{subfigure}
    \caption{The left figure shows the path coverage for the cell-specific anelastic attenuation in the CA subset of NGAWest2. The right figure shows the number of paths per cell. }
    \label{fig:cells_info}
\end{figure}

The cell-specific anelastic attenuation is modeled by $\Delta \vec{c}_{ca,p} \cdot \Delta \vec{R}$ where $\vec{c}_{ca,p}$ is vector containing the attenuation coefficients of all the cells. 
In this GMM, $\vec{c}_{ca,p}$ is modeled so that, in areas away from past paths it reverts to $c_7$ which is the anelastic attenuation coefficient in BA18, making the anelastic attenuation of the non-ergodic model equal to the anelastic attenuation of BA18 ($\vec{c}_{ca,p} \cdot \Delta \vec{R} = c_7 R_{rup}$); while in areas that are covered by the paths in NGAWest2 dataset, $\vec{c}_{ca,p}$ deviates from $c_7$ to capture the regional attenuation. 
Figure \ref{fig:cells_info} shows the cells and the path coverage in the selected subset of the NGA-West2 dataset, as well as, the number of paths per cell. 

\subsubsection{Site Scaling} \label{sec:gmm_site}

The functional form of the $f_S$ model is:

\begin{equation}
    f_S = c_8 \ln \left( \frac{\min(V_{s30}, 1000)}{1000} \right) + f_{NL} + \delta c_{1a,s} (\vec{x}_{s}) + \delta c_{1b,s}
\end{equation}
\noindent The ergodic components of the site term are: $c_8$ which controls the $V_{S30}$ scaling of the ground motion, and $f_{NL}$ which is the non-linear site amplification term. 
The non-ergodic effects related to the site are expressed by the $\delta c_{1a,s}$ and $\delta c_{1b,s}$ coefficients. 
The station constant, $\delta c_{1a,s}$, which has a finite correlation length, describes the broader adjustments to the backbone model to express the regional site effects. 
$\delta c_{1b,s}$ has a zero correlation length and acts on top of $\delta c_{1a,s}$ to describe the site specific adjustments.
Coefficients with a finite correlation length vary continuously across the domain of interest, whereas coefficients with zero correlation length vary independently from location to location. 

The remaining terms $f_{ztor}$, $f_{NM}$ and  $f_{Z1}$ where kept as they are in the ergodic $GMM$.  
 
\subsection{Formulation of spatially varying coefficient model} \label{sec:mod_stats}

The non-ergodic terms, cell-specific coefficients, and aleatory terms, hereafter collectively called model parameters ($\vec{\theta}$), were estimated by describing the GMM as a hierarchical Bayesian model using the computer software STAN \citep{Stan}. 
In Bayesian statistics, the posterior distribution of the parameters is proportional to the likelihood times the prior distribution of the parameters:

\begin{equation}
    p(\vec{\theta}|y,x)  \propto \mathcal{L}(\vec{\theta}) p(\vec{\theta})
\end{equation}
\noindent The prior distributions are the distributions that the model parameters are assumed to follow in the absence of data; the likelihood function, in general terms, is the probability of observing the data given the model parameters; and the posterior distributions are the model-parameter distributions informed by the data.

The likelihood can be estimated from the density function of the ground motion:

\begin{equation}
    \mathcal{L}(\vec{\theta}) = pdf(\vec{y}|f(x,\vec{\theta}),  \phi_0^2 + \tau^2_0)
\end{equation}

where $f(x,\vec{\theta})$ is the functional form for the median non-ergodic ground-motion:

\begin{equation}
\begin{aligned}
    f(x,\vec{\theta}) &= (f_{erg}(M,R_{rup},V_{S30},...) - c_7 R_{rup}) \\
    &+ \delta c_0 + \delta c_{0,e} + \delta c_{1,e} + \delta c_{1a,s} + \delta_{1b,s} + \vec{c}_{ca,p} \cdot  \Delta \vec{R} 
\end{aligned}
\end{equation}

\noindent It is equal to the ergodic backbone model without the effect of anelastic attenuation $f_{erg}(M,R_{rup},V_{S30},...) - c_7 R_{rup})$, plus the non-ergodic spatially varying constants that have been described the previous section ($\delta c_i$), and the cell-specific anelastic attenuation $\vec{c}_{ca,p} \cdot  \Delta \vec{R}$.

The model is called hierarchical because the prior distributions are defined in multiple levels.
At the lower level, $\vec{\theta}$ follow some prior distributions, which are defined in terms of a different set of parameters, hereafter called hyper-parameters $\vec{\theta}_{hyp}$, which, in turn, either follow some other prior distributions, or they are fixed.
In this study, the non-ergodic ergodic regression was performed in two phases: in the first phase, which included a smaller number of frequencies, $\vec{\theta}_{hyp}$ were defined by their own prior distributions that are described later in this section, whereas, in the second phase, most of $\vec{\theta}_{hyp}$ were fixed to their smoothed values, estimated from first phase, and the remaining hyperparameters were left free to follow the same prior distributions as in the first phase.
The main reasons for fixing $\vec{\theta}_{hyp}$ in the second phase were to ensure that there are no abrupt changes in the non-ergodic terms between frequencies, to constrain the model to a more physical behavior, and to reduce the computational cost of the second phase which included more frequencies.
Table \ref{tab:theta} summarizes the parameters that were classified as $\vec{\theta}$ and $\vec{\theta}_{hyp}$; the parameters composing $\vec{\theta}$ have been defined in Section \ref{sec:gmm_fun}, the hyper-parameters composing $\vec{\theta}_{hyp}$ have defined later in this section.
Table \ref{tab:free_thetahyp} summarizes the $\vec{\theta}_{hyp}$ that were free at each phase. 
If $\vec{\theta}_{hyp}$ is free, the prior distribution of a model parameter, $\theta_i$, can be explicitly defined in terms of $\vec{\theta}_{hyp}$ as follows:

\begin{equation}
    p(\theta_i) = f(\theta_i) d \theta_i = \left( \int f(\theta_i | \vec{\theta}_{hyp}) f(\vec{\theta}_{hyp}) ~d \vec{\theta}_{hyp} \right) ~d \theta_i
\end{equation}

\begin{table}
	\caption {Summary of model parameters and hyper-parameters}
	\centering
	\label{tab:theta}
	\begin{tabular}{l  c  l} 
	    \hline\noalign{\smallskip}
	    Group Name & Group Notation & \multicolumn{1}{c}{Components} \\
	    \noalign{\smallskip}\hline\noalign{\smallskip}
	    \multirow{2}{*}{Model parameters} &  \multirow{2}{*}{$\vec{\theta}$} & 
	    $\delta c_{0}$, $\delta c_{0,e}$, $\delta c_{1,e}$, $\delta c_{1a,s}$, $\delta c_{1b,s}$, \\
	    & & 
	    $c_{ca,p}$, $\delta WS^0_{e,s}$, $\delta B^0_e$  \\
	    \multirow{2}{*}{Model hyperparameters} & \multirow{2}{*}{$\vec{\theta}_{hyp}$} & 
	    $\ell_{1,e}$, $\omega_{1,e}$, $\ell_{1a,s}$, $\omega_{1b,s}$, $\omega_{1b,s}$, \\
	    & & $\ell_{ca1,p}$, $\omega_{ca1,p}$, $\omega_{ca2,p}$, $\phi_0$, $\tau_0$ \\
	    \hline\noalign{\smallskip}
	 \end{tabular}
\end{table}

\begin{table}
	\caption {Free hyper-parameters at each regression phase }
	\centering
	\label{tab:free_thetahyp}
	\begin{tabular}{c l}
	    \hline\noalign{\smallskip}
	    Phase & \multicolumn{1}{c}{Free hyper-parameters} \\ 
	    \noalign{\smallskip}\hline\noalign{\smallskip}
	    1       & $\ell_{1,e}$, $\omega_{1,e}$, $\ell_{1a,s}$, $\omega_{1a,s}$, $\omega_{1b,s}$, $\ell_{ca1,p}$, $\omega_{ca1,p}$, $\omega_{ca2,p}$, $\phi_0$, $\tau_0$ \\
	    2       & $\phi_0$, $\tau_0$ \\
	    \hline\noalign{\smallskip}
	 \end{tabular}
\end{table}

More specifically, the $\delta c_0$ constant has a normal prior distribution with a zero mean and a $0.1$ standard deviation:

\begin{equation}
    \delta c_0 \sim \mathcal{N}(0, 0.1) 
\end{equation}
\noindent The mean is set to zero because in absence of data, there should be no shift between the ergodic and non-ergodic GMM.
The standard deviation is small because any constant shift informed by the regional data or the re-weighting of the residuals is expected to be small. 

For earthquakes with magnitudes less than $M~5$ and frequencies less than $5~Hz$, $\delta c_{0,e}$ follow a normal prior distribution with a zero mean and a $0.2$ standard deviation. 
Preliminary analyses, which did not include $\delta c_{0,e}$, had a $0.2$ to $0.4$ regional difference in $\delta c_{1,e}$ between northern and southern California, which have a $16$ and $3 \%$ probability of being exceeded with the selected standard deviation. 
Therefore, the posterior distribution of $\delta c_{0,e}$ will deviate from zero to reach a similar range only if there is significant support by the data; otherwise,  $\delta c_{0,e}$ will stay close to zero implying no systematic difference between northern and southern CA at small magnitude events. 

\begin{equation}
    \delta c_{0,e} \sim \left\{ \begin{array}{lllll}
        0                   & for& M > 5 &or&  f> 5 Hz \\
        \mathcal{N}(0, 0.2) & for& M < 5 &and& f < 5 Hz
    \end{array} 
    \right. 
\end{equation}

The non-ergodic constants $\delta c_{1,e}(\vec{x}_{e})$ and  $\delta c_{1a,s}(\vec{x}_{s})$ follow multivariate normal prior distributions with zero mean and Matern (negative exponential) covariance functions ($\kappa$).
$\kappa$ imposes the spatial correlation on $\delta c_{1,e}$ and $\delta c_{1a,s}$: it ensures that the values of $\delta c_{1,e}$ and $\delta c_{1a,s}$ will be similar for earthquakes or sites in close proximity and that $\delta c_{1,e}$ and $\delta c_{1a,s}$ would vary continuously from location to location.

\begin{equation}
\begin{aligned}
    \delta c_{1,e}  &\sim \mathcal{N}(0, \kappa_{1,e} ) \\
    \delta c_{1a,s} &\sim \mathcal{N}(0, \kappa_{1a,s} )
\end{aligned}
\end{equation}

The covariance function between a pair of earthquakes for $\kappa(\vec{x}_{e}, \vec{x}'_{e})$ or between a pair of stations for $\kappa(\vec{x}_{s}, \vec{x}'_{s})$ is defined in equation \eqref{eq:kernel_const_dcoeffs}; $\vec{x}$ and $\vec{x}'$ are the coordinates of the two earthquakes or sites depending on the coefficient, $\omega_i$ is the standard deviation, and $\ell_i$ is the correlation length. 
$\omega_i$ controls the variability of the non-ergodic coefficients, that is, how much the values of the coefficients could vary between locations that are far from each other.
$\ell_i$ governs the length scale of the spatial variation of $\delta c_i$; increasing $\ell_i$ makes $\delta c_i$ to vary more gradually with distance.

\begin{equation} \label{eq:kernel_const_dcoeffs}
\begin{aligned}
    \kappa_{1,e}(\vec{x}_e, \vec{x}'_e)  &= \omega_{1,e}^2  ~ e^{-\frac{ \lVert \vec{x}_e - \vec{x}'_e \lVert }{ \ell_{1,e}}  } \\
    \kappa_{1a,s}(\vec{x}_s, \vec{x}'_s) &= \omega_{1a,s}^2 ~ e^{-\frac{ \lVert \vec{x}_s - \vec{x}'_s \lVert }{ \ell_{1a,s}} } \\
\end{aligned}
\end{equation}

Both $\ell_{1,e}$ and $\ell_{1a,s}$ have inverse gamma prior distributions with distribution parameters $\alpha$ and $\beta$ equal to $2$ and $50$ which corresponds to a mode and mean of $16.7$ and $50~km$, respectively.

\begin{equation} 
\begin{aligned}
    \ell_{1,e}  &\sim InvGamma(2.0,50.0) \\
    \ell_{1a,s} &\sim InvGamma(2.0,50.0) \\
\end{aligned}
\end{equation}

\noindent Inverse gamma distributions are defined only for positive real numbers which is a desirable property for the prior distributions of the correlation lengths as negative correlation lengths do not have any physical meaning.
The $\ell_{1,e}$ and $\ell_{1a,s}$ correlation lengths are expected to be around $10$ to $50~km$, where the inverse gamma distribution has most of its mass, but larger values are also possible, if they are supported by the data, due to its exponential tail. 

The prior distribution of $\omega_{1,e}$ and $\omega_{1a,s}$ is an exponential distribution with a rate of $20$

\begin{equation} 
\begin{aligned}
    \omega_{1,e}  &\sim exp(20) \\
    \omega_{1a,s} &\sim exp(20) \\
\end{aligned}
\end{equation}

\noindent These prior distributions were chosen for $\omega_{1,e}$ and $\omega_{1a,s}$ to penalize unnecessary model complexity \citep{Simpson2017}.
A $\omega_i$ equal to zero implies no variability for $\delta c_i$, meaning that there are no systematic effects related to that parameter.
In an exponential distribution, most of the mass is near zero, which allows  $\delta c_i$ to deviate from zero to capture systematic effects related to that parameter only if there is significant support by the ground-motion observations.
For the same reason, exponential prior distributions were used in \cite{Kuehn2020b} to model the standard deviations of the regional terms in the KBCG20 partially non-ergodic subduction-zone GMM.

The site-specific adjustment, $\delta c_{1b,s}$ follows a normal distribution with a zero mean and a $\omega_{1b,s}$ standard deviation:

\begin{equation}
    \delta c_{1b,s} \sim \mathcal{N}(0, \omega_{1b,s})
\end{equation}

$\delta c_{1b,s}$ is a function of the site location, the same adjustment is applied to all ground motions recorded at the same station.

The prior distribution for $\omega_{1b,s}$ is a log-normal distribution with a logmean of $-0.8$ and a standard deviation of $0.3$ natural log units:

\begin{equation}
    \omega_{1b,s} \sim \mathcal{LN}(-0.8, 0.3)
\end{equation}

\noindent This prior distribution has a median value of 0.45, and a $16^{th}$ and $84^{th}$ percentile of $0.33$ and $0.6$, respectively.
\cite{Bayless2019a} found $\phi_{S2S}$ to range from $0.4$ to $0.6$ which is consistent with the prior distribution for $\omega_{1a,s}$.

The prior distribution of $\vec{c}_{ca,p}$ is a multivariate normal distribution with an upper truncation limit at zero:

\begin{equation}
    \vec{c}_{ca,p} \sim \mathcal{N}(\mu_{ca,p}, \kappa_{ca,p}) \mathcal{T}(,0)
\end{equation}

\noindent where $\mu_{ca,p}$ is the mean of the distribution, and $\kappa_{ca,p}$ is the covariance function.
To ensure the physical extrapolation of the GMM, $\vec{c}_{ca,p}$ is limited to be less or equal to zero, which is satisfied by setting the upper limit of the normal distribution at zero ($T(,0)$).
Two key differences from the \cite{Abrahamson2019} approach when modeling the cells are the different mean and the different covariance function of the prior distribution.
In this model, the mean of the prior is equal to the value of the anelastic-attenuation coefficient in \cite{Bayless2019a} ($\mu_{ca,p} = c_{7~BA18}$); thus, in areas with sparse data, the non-ergodic attenuation goes back to the ergodic attenuation to ensure reasonable extrapolation at large distances. 
This decision was made because the local data may not be sufficient to estimate both the median shift and the spatial variability of the anelastic attenuation. 
The covariance function (equation \eqref{eq:cA_cov}) is the sum of a Matern kernel scaled by $\omega_{ca1,p}^2$ and a diagonal kernel scaled by $\omega_{ca2,p}^2$.
The Matern kernel controls the underlining continuous variation of anelastic attenuation over large areas, whereas, the diagonal kernel allows for some independence in the attenuation from cell to cell. $\omega_{ca1,p}$ controls the size of the underling variability of $c_{ca,p}$ over large distances, the correlation length $\ell_{ca1,p}$ controls how fast the underling component of $c_{ca,p}$ varies with distance, and $\omega_{ca2,p}$ controls the size of the independent variability.
The location of each cell is defined in terms of its midpoint coordinates, $\vec{c_c}$; the distance between the cells is calculated by the L2 norm of the vector between midpoints of the cells. 

\begin{equation} \label{eq:cA_cov}
    \kappa_{ca,p}(\vec{x}_c, \vec{x}'_c) = \omega_{ca1,p}^2 ~ e^{ -\frac{ \lVert \vec{x}_c - \vec{x}'_c \lVert }{ \ell_{ca1,p}} } 
                                      + \omega_{ca2,p}^2 \delta( \lVert \vec{x}_c - \vec{x}'_c \lVert )
\end{equation}

$\ell_{ca1,p}$ has an inverse gamma distribution with the same parameters as the prior distributions for $\ell_{1e}$ and $\ell_{1a,s}$.
$\omega_{ca1,p}$ and $\omega_{ca2,p}$ have an exponential prior distribution with the same parameters as the prior distributions for $\omega_{1,e}$ and $\omega_{1a,s}$.

\begin{equation} 
\begin{aligned}
    \ell_{ca,p}    &\sim InvGamma(2.0,50.0) \\
    \omega_{ca1,p} &\sim exp(20) \\
    \omega_{ca2,p} &\sim exp(20) \\
\end{aligned}
\end{equation}

The non-ergodic within-event residuals, $\delta W^0_{e,s}$, follow a normal distribution with a zero mean and $\phi_0$ standard deviation. 

\begin{equation}
    \delta W^0_{e,s} \sim \mathcal{N}(0, \phi_0^2)
\end{equation}

The prior distribution of $\phi_0$ is a log-normal distribution with a logmean of $-1.3$ and a standard deviation of $0.3$ natural log units (equation \eqref{eq:phi0_prior}).
This set of parameters leads to a prior mean of $0.27$ and a $16^{th}$ to $84^{th}$ percentile range of $0.2$ to $0.37$. 
BA18 is an ergodic model, and so an estimate of $\phi_0$ is not available to inform the prior distribution of $\phi_0$ of this model.
However, $\phi_{SS}$, which is available in BA18, is about 0.4 for most frequencies, and because $\phi_{0}$ is smaller than $\phi_{SS}$ by definition \citep{AlAtik2010}, the range of the prior distribution is reasonable. 

\begin{equation} \label{eq:phi0_prior}
    \phi_{0} \sim \mathcal{LN}(-1.3, 0.3)
\end{equation}

The non-ergodic between-event residuals, $\delta B^0_{e}$, follow a normal distribution:
\begin{equation} \label{eq:dBo_dist}
    \delta B^0_e \sim \mathcal{N}(0, \tau_0 )
\end{equation} 
\noindent with a zero mean and $\tau_0$ standard deviation.
$\delta B^0_e$ is a function of the earthquake id, $e$; that is, the same $\delta B^0_e$ is applied to all recordings of the same earthquake.

The prior distribution of $\tau_0$ is a log-normal distribution with a $-1.0$ logmean and $0.3$ log-standard deviation. 

\begin{equation}
    \tau_0 \sim \mathcal{LN}(-1.0,0.3)
\end{equation}

The $\tau_0$ distribution parameters are judged to be reasonable because the mean and $16^{th}$ and $84^{th}$ percentiles ($0.38$, $0.27$ and $0.50$) are in agreement with other non-ergodic studies where the total non-ergodic standard deviation ($\sqrt{\phi_0^2 + \tau_0^2}$) ranges from $0.36$ to $0.55$.

\subsection{Predictive distributions of coefficients at new locations} \label{sec:predict_dist}

The non-ergodic coefficients can be estimated at new locations ($\vec{x}^*$) by conditioning them on the non-ergodic coefficients at the existing locations ($\vec{x}$); that is, the location of stations or past events depending on the coefficient. 
Since all non-ergodic coefficients follow multivariate normal distributions, for known values of the non-ergodic coefficients ($\vec{\delta c_i}$) at the $\vec{x}$, the non-ergodic coefficients at $\vec{x}^*$ also follow multivariate normal distributions with the mean and covariance matrix \citep{Rasmussen2006, Landwehr2016}: 

\begin{equation} \label{eq:cond_mean_fixed}
    \vec{\mu}_{\delta c^*_i | \delta c_i } = \mathbf{k}_i^\intercal \mathbf{K}_i^{-1} \vec{ \delta c_i }
\end{equation}
\begin{equation} \label{eq:cond_cov_fixed}
    \mathbf{\Psi}_{\delta c_i^* | \delta c_i } =  \mathbf{K^*}_i - \mathbf{k}_i^\intercal \mathbf{K}_i^{-1}  \mathbf{k}_i
\end{equation}

\noindent where $\vec{\mu}_{\delta c^*_i | \delta c_i }$ mean of non-ergodic ergodic coefficients at $\vec{x}'$ conditioned on $\vec{\delta c_i}$, $\mathbf{\Psi}_{\delta c^*_i | \delta c_i }$ is the covariance of non-ergodic coefficients at $\vec{x}'$ conditioned on $\vec{\delta c_i}$, $\mathbf{K}$ is the covariance between the non-ergodic coefficients at the existing locations ($\mathbf{K}_i = \kappa_i(\vec{x},\vec{x})$), $\mathbf{k}$ is the covariance between the non-ergodic coefficients at the existing and new locations ($\mathbf{k}_i = \kappa_i(\vec{x},\vec{x}^*)$), and $\mathbf{K^*}$ is the covariance between the non ergodic coefficients at the new locations ($\mathbf{K^*}_i = \kappa_i(\vec{x}^*,\vec{x}^*)$). 

However, the non-ergodic coefficients at $\vec{x}$ are not perfectly known.
There is some epistemic uncertainty associated with $\vec{\delta c}_i$ which is quantified by their posterior distribution. 
To simplify the calculations and obtain a closed-form solution, $\vec{\delta c}_i$ is assumed to follow a multivariate normal posterior distribution ($\vec{\delta c}_i \sim \mathcal{N}( \vec{\mu}_{\delta c_i}, \mathbf{\Psi}_{\delta c_i} )$); $\vec{\mu}_{\delta c_i}$ is the posterior mean, and $\mathbf{\Psi}_{\delta c_i}$ is a diagonal matrix with the posterior variances across the diagonal.
We consider this assumption to be reasonable because all $\vec{\delta c}_i$ have multivariate normal prior distributions. 
If all of the hyper-parameters were fixed, the posterior distributions of $\vec{\delta c}_i$ would indeed be multivariate normal distributions, but because some of the hyper-parameters are free, the posterior distributions of $\vec{\delta c}_i$ may slightly deviate from the assumption. 
The epistemic uncertainty of $\delta c_i$ can be included in the $\delta c_i^*$ predictions by using the marginal distribution of $\delta c_i^*$:

\begin{equation} 
    p(\delta c^*_i)  = \int p(\delta c^*_i |\delta c_i) p( \delta c_i )~ d \delta c_i 
\end{equation}

Due to the previous assumption, $p(\delta c^*_i)$ is also a multivariate normal distribution with mean and covariance matrix given in Equations \eqref{eq:cond_mean} and \eqref{eq:cond_cov}, respectively \citep{Bishop2006}.

\begin{equation} \label{eq:cond_mean}
    \vec{\mu}_{\delta c^*_i} = \mathbf{k}_i^\intercal \mathbf{K}_i^{-1} \vec{ \mu  }_{\delta c_i}
\end{equation}
\begin{equation} \label{eq:cond_cov}
    \mathbf{\Psi}_{\delta c_i^*} =  \mathbf{K^*}_i - \mathbf{k}_i^\intercal \mathbf{K}_i^{-1}  \mathbf{k}_i + 
        \mathbf{k}_i^\intercal \mathbf{K}_i^{-1}  \mathbf{\Psi}_{\delta c_i }  (\mathbf{k}_i^\intercal \mathbf{K}_i^{-1})^\intercal 
\end{equation}

\subsection{Inter-frequency Correlation} \label{sec:ifreq_model}
The main motivation behind the development of this non-ergodic EAS GMM is to use it with RVT to create an equivalent non-ergodic $PSA$ GMM; in doing this, it is important to capture the inter-frequency correlation of the non-ergodic terms, otherwise, as it was demonstrated by \cite{Bayless2018}, the variability of the $PSA$ values is underestimated.  
The correlation coefficient ($\rho$) is a measure of the linear relationship of two random variables $X_1$ and $X_2$. A $\rho$ that is equal to one implies that $X_2$ can be perfectly defined as a linear function of $X_1$, and vice versa; a zero $\rho$ implies that the two random variables are uncorrelated. 
In ground-motion studies, the inter-frequency correlation coefficient is a measure of the width of the peaks and troughs of a $PSA$ or $EAS$ spectrum: the stronger the correlation of the amplitudes between frequencies, the wider the peaks and troughs of the spectra will be. 

The correlation coefficient for a non-ergodic term $\delta c_i$, at frequencies $f_1$ and $f_2$, is defined as:

\begin{equation}
    \rho_{\delta c_i}(f_1,f_2) = \frac{ cov \left( \delta c_i(f_1), \delta c_i(f_2) \right) }{ \sigma_{\delta c_i(f_1)} \sigma_{\delta c_i(f_2)} }
\end{equation}
\noindent where $cov$ is the covariance of $\delta c_i$ at the two frequencies, and $\sigma_i$ is the standard deviation of $\delta c_i$.
$\rho$ can be determined from the data using the maximum likelihood estimator \citep{Kutner2005}:
\begin{equation}
    \rho_{\delta c_i}(f_1,f_2) = \frac{\sum_{j=1}^{n} \left( \delta c_{i,j}(f_1)-\overline{\delta c_{i}}(f_1) \right) 
                                                    \left( \delta c_{i,j}(f_2) - \overline{\delta c_{i}}(f_2) \right)  }
                                         {\sqrt{\sum_{j=1}^{n} \left( \delta c_{i,j}(f_1)-\overline{\delta c_{i}}(f_1) \right)^2}
                                          \sqrt{\sum_{j=1}^{n} \left( \delta c_{i,j}(f_2)-\overline{\delta c_{i}}(f_2) \right)^2} }
\end{equation}
\noindent where $n$ is the number of observations, $\delta c_{i,j}$ is the $j^{th}$ sample of $\delta c_i$, and $\overline{\delta c_{i}}$ is the mean value of $\delta c_i$.
For a large number of samples ($n > 25$), $\rho$ can be transformed into a random variable $z$ that follows a normal distribution with equation \eqref{eq:rho_trans} \citep{Kutner2005}; the standard deviation of $z$ is given in equation \eqref{eq:rho_trans_std}.

\begin{equation} \label{eq:rho_trans}
    z = \tanh^{-1}(  \rho ) = \frac{1}{2} ~ \ln\left( \frac{1+\rho}{1-\rho} \right)
\end{equation}

\begin{equation} \label{eq:rho_trans_std}
    \sigma(z) = \sqrt{ \frac{1}{n-3} }
\end{equation}

The same functional form that was used to model the correlation of the total EAS residuals in \cite{Bayless2019} was used here to fit the empirical correlations of the non ergodic terms:

\begin{equation} \label{eq:rho_model}
    \rho(f_r) = \left\{ \begin{array}{cll}
                    1                                               & for & f_r = 0  \\
                    \tanh \left( A e^{B f_r} + C e^{D f_r} \right)  & for & f_r \neq 0   \\
    \end{array} 
    \right. 
\end{equation}

\begin{equation}
    f_r = \left| \ln \left( \frac{f_1}{f_2} \right) \right|
\end{equation}

\noindent $A$, $B$, $C$, $D$ are the model parameters, and $f_r$ is the absolute value of the natural log of the ratio of the two frequencies. 
This functional form allows for a two-term exponential decay as a function of the logarithm of frequency; this behavior is required because both the correlation of the total residuals in \cite{Bayless2019} and the correlation of the epistemic uncertainty terms presented in section \ref{sec:results} exhibit a steep decay at frequencies near the conditioning frequency which then flattens at frequencies that are further from the conditioning frequency. 
The model parameters were estimated with a non-linear least-squares regression on $z$ using the MINPACK.LM package \citep{minpacklm} in the statistical software R \citep{R}; $\sigma(z)$ was used as weights in the least-squares regression emphasising the fit to the higher correlation values which have more samples.

In this study, one difference from the \cite{Bayless2019} is it that the inter-frequency correlation of all epistemic uncertainty terms was modeled as frequency independent (i.e. $A$, $B$, $C$, $D$ are constants).
This was done, because at it can be seen in section \ref{sec:results}, $\delta c_{1a,s}$ and  $\delta c_{1b,s}$, which are the biggest contributors to the total non-ergodic effects, have an almost frequency independent inter-frequency correlation.
$\vec{c}_{ca,p}$ show the most noticeable frequency dependence in inter-frequency correlation, but it becomes more stable at intermediate and large frequencies which is the frequency range where it has the biggest impact. 
The assumption of frequency independence should be re-examined in future studies with a larger dataset. 

\section{Results} \label{sec:results}

\subsection{Hyperparameters} \label{sec:hyp}

\begin{figure}
    \centering
    \begin{subfigure}[t]{0.28\textwidth} 
        \caption{} \label{fig:smoothed_rhoX1a}  
        \includegraphics[width = .95\textwidth]{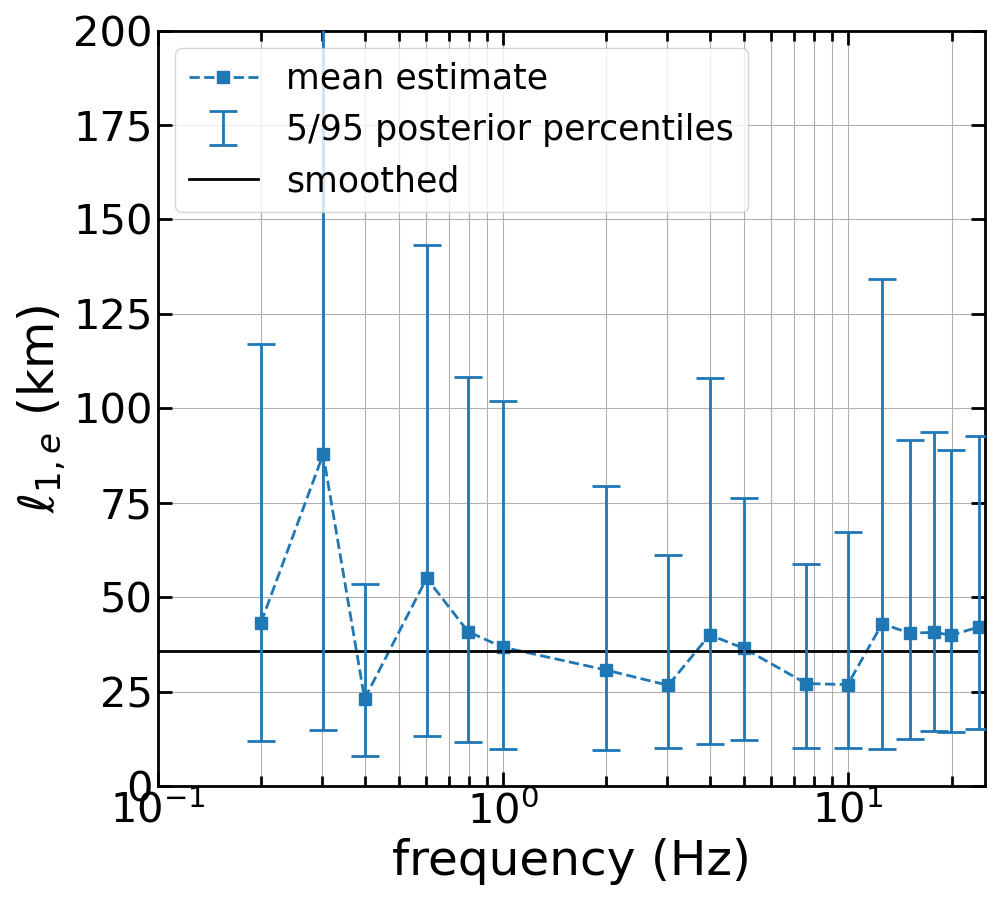}
    \end{subfigure}
    \begin{subfigure}[t]{0.28\textwidth}
        \caption{} \label{fig:smoothed_thetaX1a} 
        \includegraphics[width = .95\textwidth]{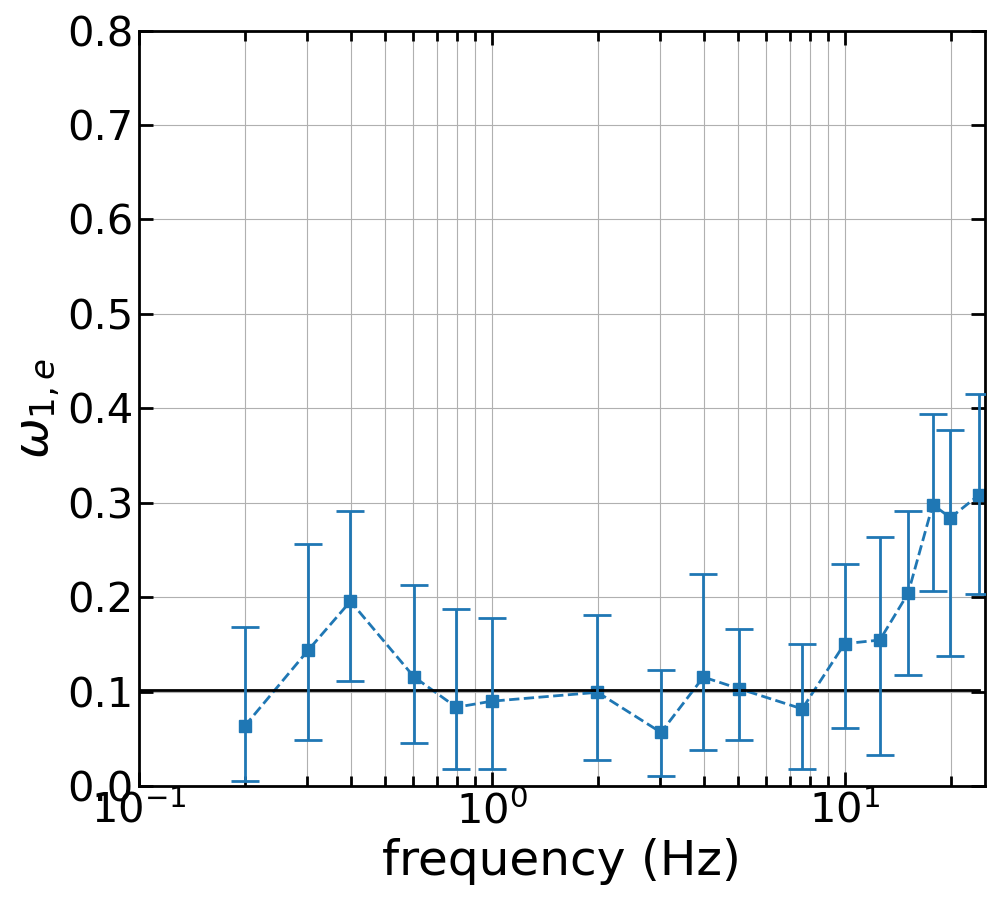}
    \end{subfigure}
    \\
    \begin{subfigure}[t]{0.28\textwidth}
        \caption{} \label{fig:smoothed_rhoX1b}  
        \includegraphics[width = .95\textwidth]{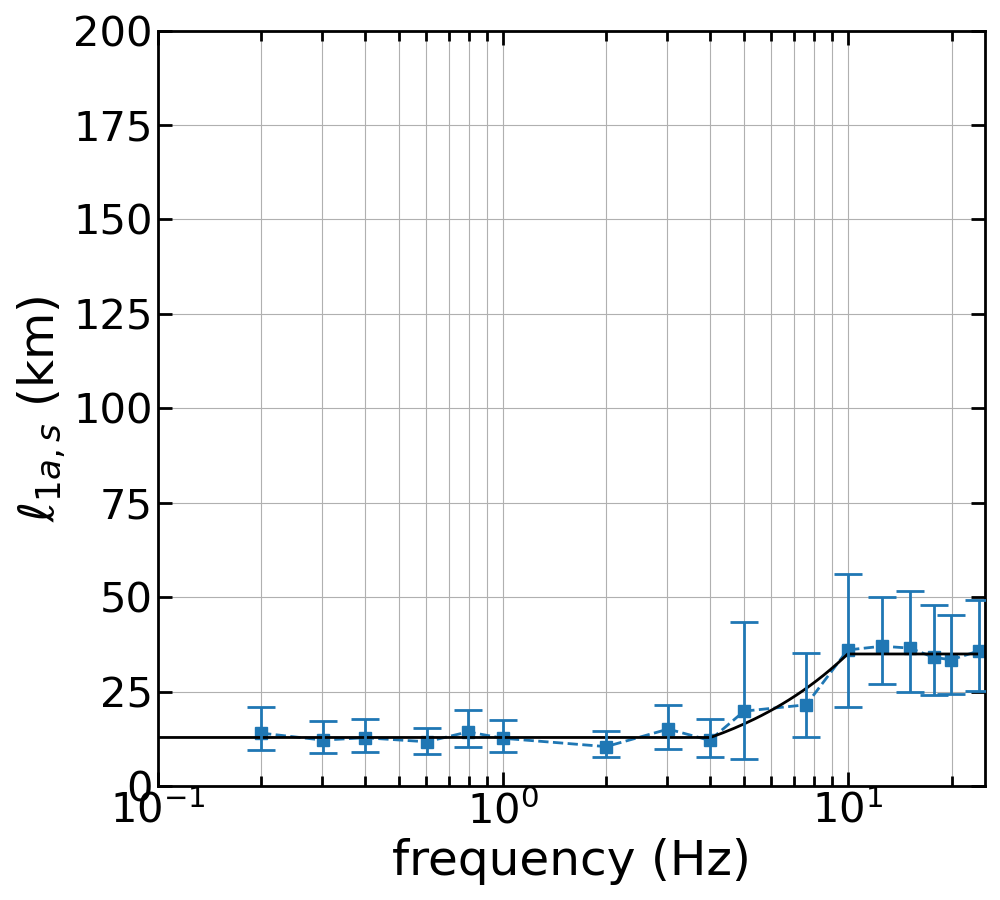}
    \end{subfigure}  
    \begin{subfigure}[t]{0.28\textwidth}
        \caption{} \label{fig:smoothed_thetaX1b}
        \includegraphics[width = .95\textwidth]{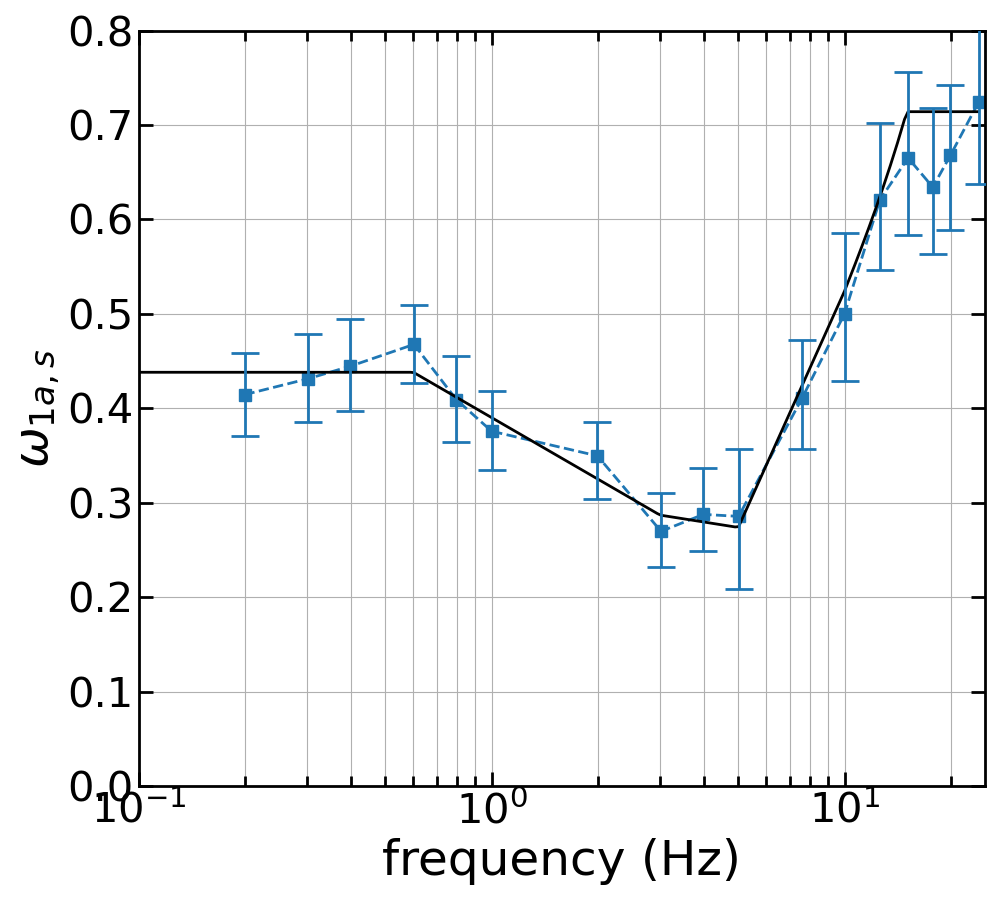}
    \end{subfigure}  
    \begin{subfigure}[t]{0.28\textwidth}
        \caption{} \label{fig:smoothed_phiSS}
        \includegraphics[width = .95\textwidth]{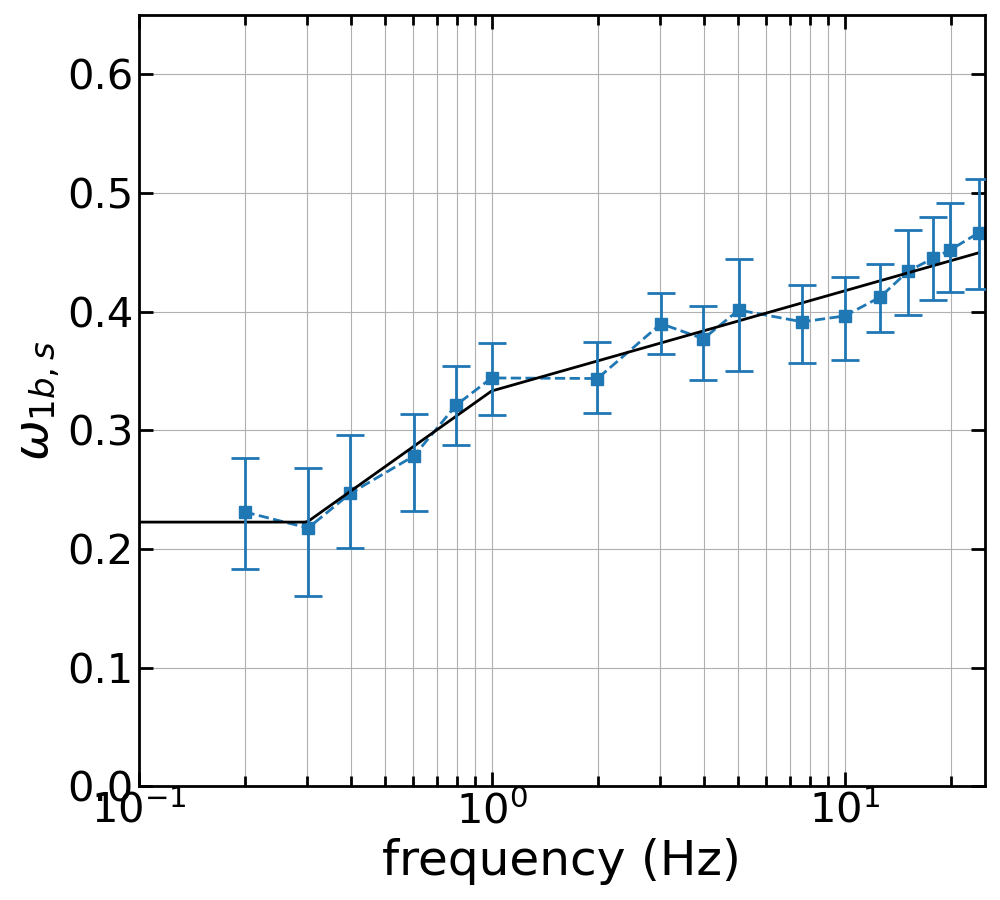}
    \end{subfigure}  
    \\
    \begin{subfigure}[t]{0.28\textwidth}
        \caption{}
        \includegraphics[width = .95\textwidth]{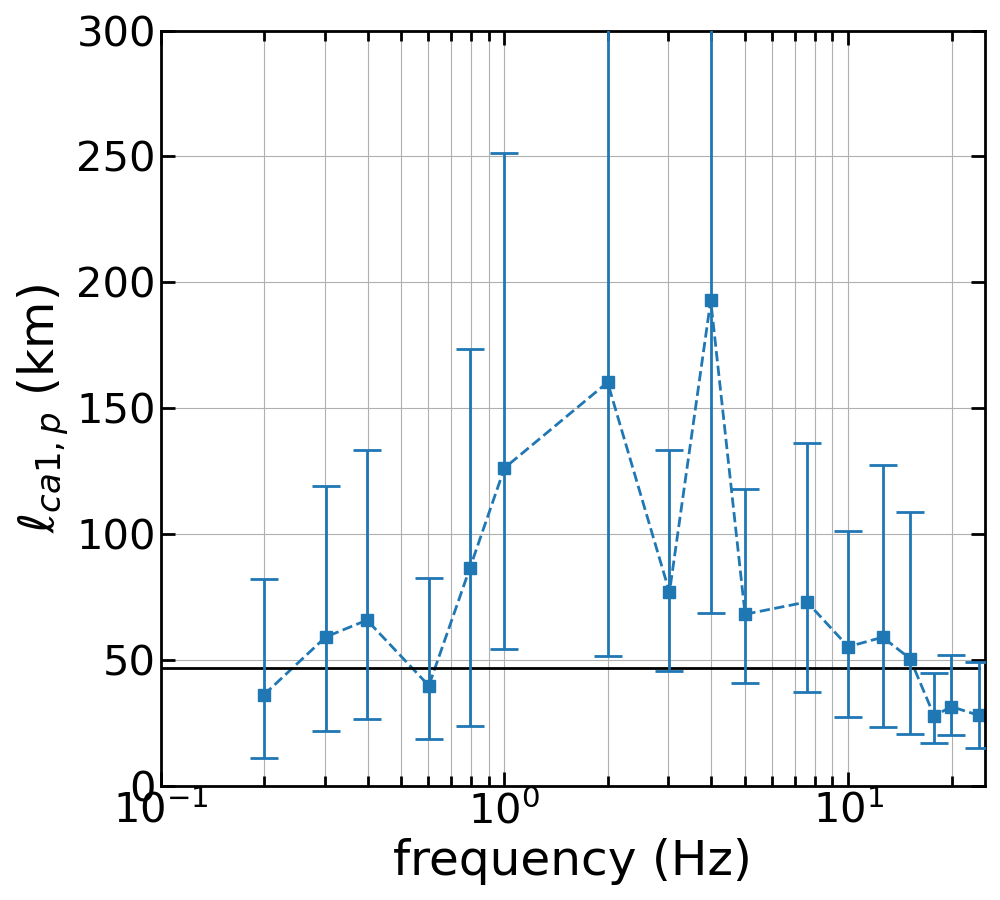}
    \end{subfigure}  
    \begin{subfigure}[t]{0.28\textwidth}
        \caption{}
        \includegraphics[width = .95\textwidth]{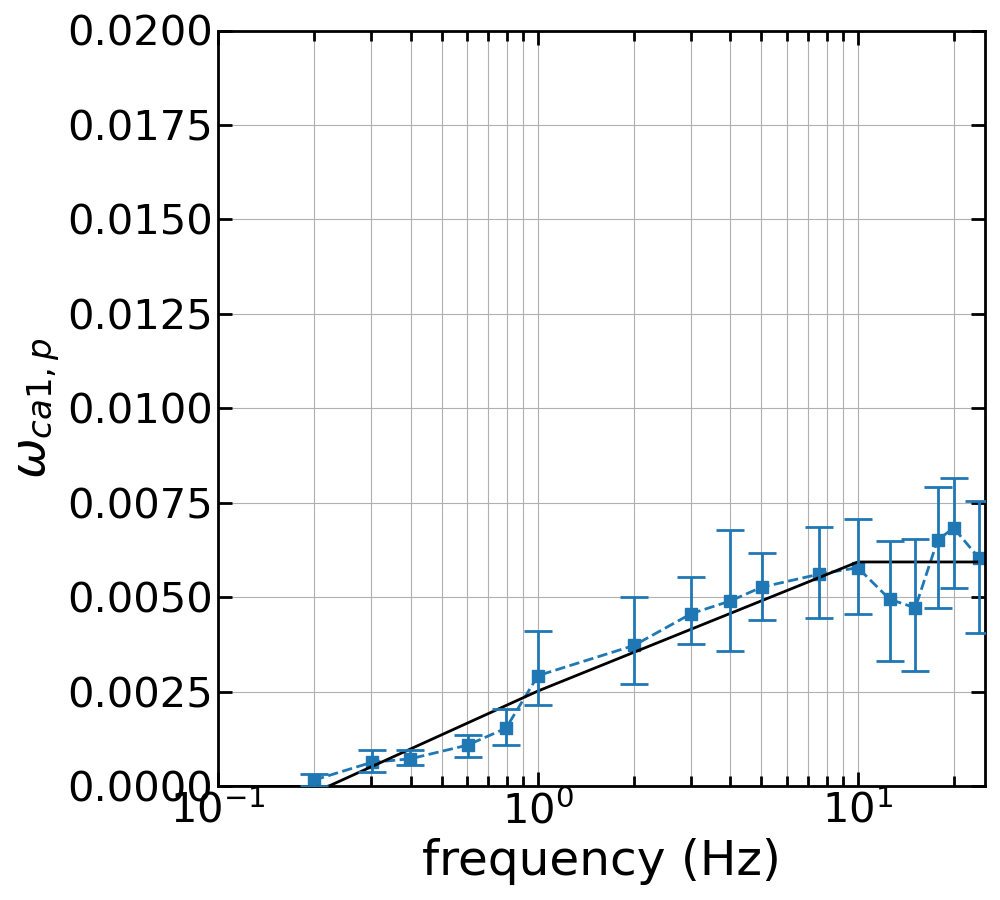}
    \end{subfigure}  
    \begin{subfigure}[t]{0.28\textwidth}
        \caption{}
        \includegraphics[width = .95\textwidth]{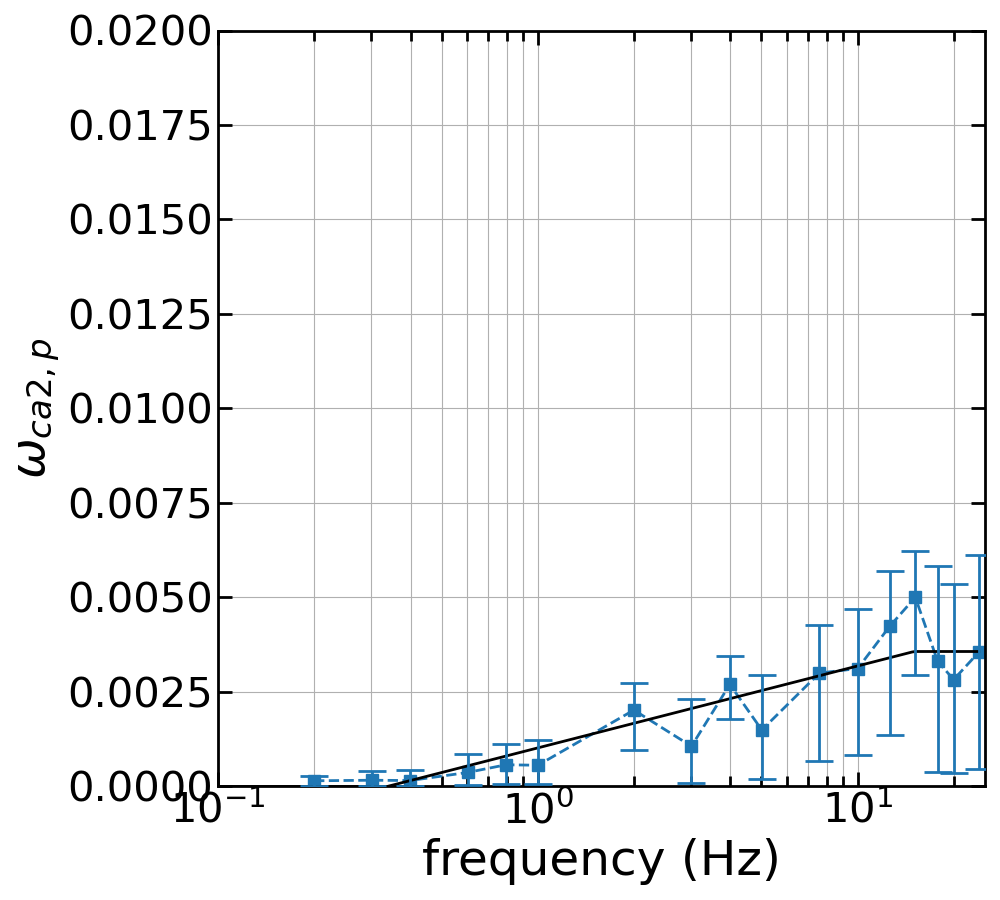}
    \end{subfigure}  
    \caption{Estimated and smoothed hyperparameters versus frequency; the circular  line depicts the mean estimate of the hyperparameters from the  original regression, 
    the vertical bars correspond to the $5/95$ percentiles of the posterior distribution, 
    and the solid line represent to the smoothed hyperparameters. 
    (a) correlation length of the source constant, $\ell_{1,e}$, 
    (b) standard deviation of source constant, $\omega_{1,e}$, 
    (d) correlation length of the site constant with finite correlation length, $\ell_{1a,s}$,
    (e) standard deviation of the site constant with finite correlation length, $\omega_{1a,s}$,
    (c) standard deviation of the site term with zero correlation length, $\omega_{1b,s}$,
    (f) correlation length of the cell-specific anelastic attenuation, $\ell_{ca1,p}$, 
    (g) standard deviation of the uncorrelated component of the cell-specific anelastic attenuation, $\omega_{ca1,p}$, and
    (h) standard deviation of the correlated component of the cell-specific anelastic attenuation, $\omega_{ca2,p}$.}
    \label{fig:smoothed_hyp}
\end{figure}

Figure \ref{fig:smoothed_hyp} presents the mean, $5^{th}$ and $95^{th}$ percentiles of the posterior distribution of the hyper-parameters of the non-ergodic terms; the proposed smoothed values are also presented in the same figure.
As mentioned in section \ref{sec:mod_stats}, the regression for this model was performed in two phases; in the first phase, all model hyperparameter were free and estimated based on the data and prior distributions, whereas in the second phase, the hyper-parameters of the non-ergodic terms were fixed to their smoothed values, and $\tau_0$ and $\phi_0$ were reestimated for the new set of smoothed hyper-parameters.

Figure \ref{fig:smoothed_rhoX1a} shows the variation of the correlation length for $\delta c_{1,e}$ with frequency.
The posterior distribution of $\ell_{1,e}$ is wide due to the small number of earthquakes.
Overall, the mean estimate of $\ell_{1,e}$ is around $40 km$ except at low frequencies where at $0.3Hz$ it jumps up to $85 km$.
Because there is no reason for $\ell_{1,e}$ to increase at low frequencies, it was fixed to the average $\ell_{1,e}$ over all frequencies, as it the simplest model.
Furthermore, there are less data at low and high frequencies making the estimates of the hyperparameters at these frequency ranges less stable.

Figure \ref{fig:smoothed_thetaX1a} shows how $\omega_{1,e}$ changes with frequency.
In this case, the posterior mean of $\omega_{1,e}$ stays constant at low and intermediate frequencies, and it exhibits an increase at high frequencies.
Similarly to $\ell_{1,e}$, there is no reason for $\omega_{1,e}$ to increase at high frequencies so it was fixed to the average value over all frequencies.
One possible cause of the apparent increase of $\omega_{1,e}$ at high frequencies is that some of the non-ergodic site effects could have been mapped into non-ergodic source effects in the regression.
It is expected that the non-ergodic site effects will increase at high frequencies because the regional differences in site amplification tend to have a larger impact on the high frequencies.
This assumption is consistent with the behavior of $\omega_{1a,s}$ and $\omega_{1b,s}$, which both show an increase with frequency. 
For this reason, the difference between the estimated and smoothed $\omega_{1,e}$ was moved to $\omega_{1a,s}$.

In smoothing $\omega_{1a,s}$, up to the frequency of $15 Hz$, a piece-wise linear model was fit to the estimated mean values; whereas beyond $15 Hz$, it was fit to the square root of the sum of squares of the estimated mean $\omega_{1a,s}$ and the difference between the estimated mean and smoothed $\omega_{1,e}$.
Minimal smoothing was applied to $\ell_{1a,s}$ and $\omega_{1b,s}$ as they show a relatively small variation between neighbouring frequencies. 

The smoothed $\ell_{ca1,p}$ was fixed to the average of the mean estimates that are less than $75 km$; this upper limit was imposed because the $\ell_{ca1,p}$ with large mean estimates also had wide posterior distributions, meaning that $\ell_{ca1,p}$ could not be reliably estimated at those frequencies.

$\omega_{ca1,p}$ and $\omega_{ca2,p}$ exhibit similar characteristics in their variation with frequency: they are very small at low frequencies, they show an approximately linear increase with the log of frequency at intermediate frequencies, and they reach a plateau at high frequencies.
This happens because the effects of anelastic attenuation are more noticeable at high frequencies, and likewise, spatial changes in the anelastic attenuation will have a larger effect on higher frequencies.
This behavior was also observed by \cite{Kuehn2019} who found that the standard deviation of the cell-specific attenuation coefficients of their non-ergodic $PSA$ GMM was smaller at long periods .

\begin{figure}
    \centering
    \begin{subfigure}[t]{0.40\textwidth} 
        \caption{} 
        \includegraphics[width = .95\textwidth]{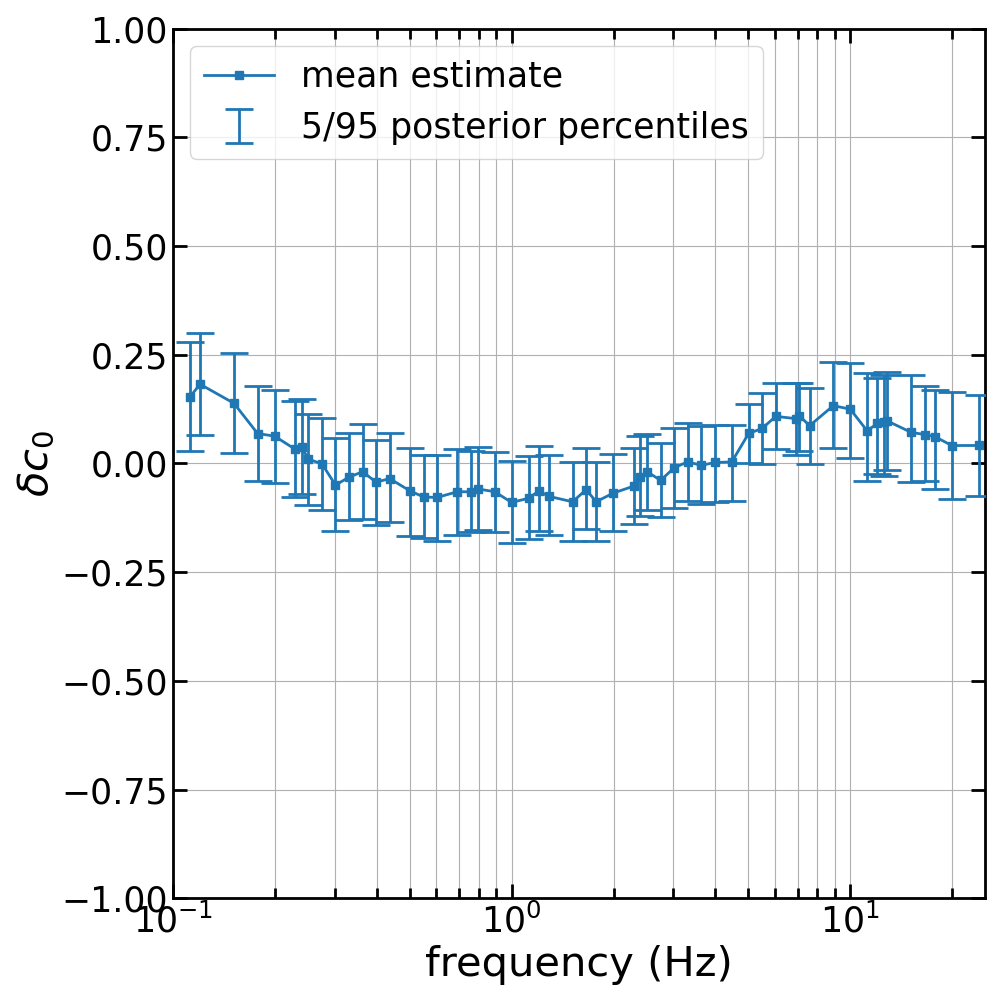}
    \end{subfigure}
    \begin{subfigure}[t]{0.40\textwidth}
        \caption{}  
        \includegraphics[width = .95\textwidth]{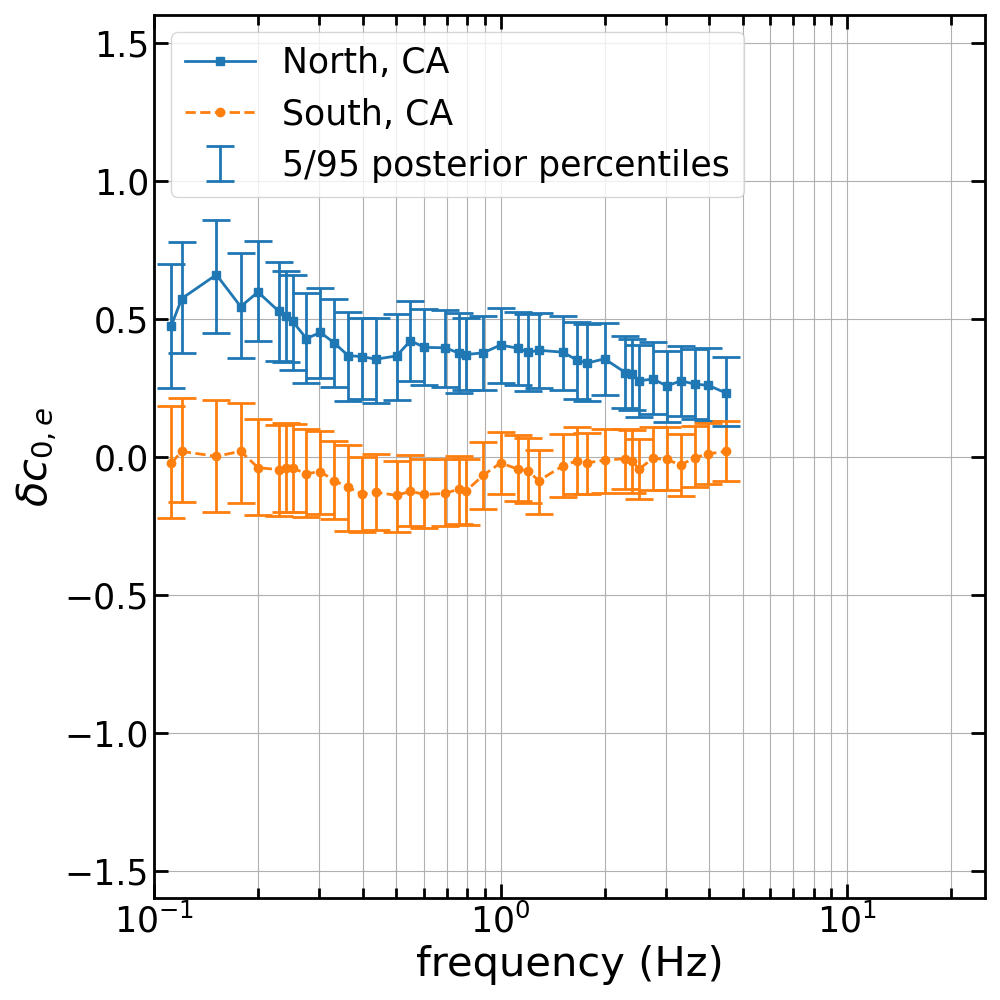}
    \end{subfigure}
    \caption{Estimated $\delta c_{0}$ and $\delta c_{0~N/S}$ versus frequency;
    the circular marker corresponds to the mean estimate, 
    the vertical bars represents the $5/95$ percentiles of the posterior distribution
    (a) constant shift, $\delta c_{0}$, 
    (b) regional shift, $\delta c_{0,e}$, the solid line with the square markers corresponds to the Northern CA, the dashed line with the circular markers corresponds to the Southern CA}
    \label{fig:con}
\end{figure}

Figure \ref{fig:con} presents the terms that were reestimated in the second step. 
The coefficient $\delta c_0$ corresponds to the shift of the non-ergodic GMM due to the reweighting of the residuals.
For most frequencies, the change in the constant from the ergodic model is less than $10 \%$. 
The regional constant $\delta c_{0,e}$, which corrects for the bias due to the differences in the small magnitude conversion, is about $0.4$ for the northern California and zero for the southern California.
In the NGAWest2 dataset, most of the data are located in southern California, so it is expected that the base model would be consistent with the southern part of the state with the main correction applied to the northern part of California.

\subsection{Spatially varying coefficients and cell-specific anelastic attenuation} \label{sec:coeffs}

\begin{figure}
    \centering
    \begin{subfigure}[t]{0.4\textwidth} 
        \caption{}
        \includegraphics[width = .95\textwidth]{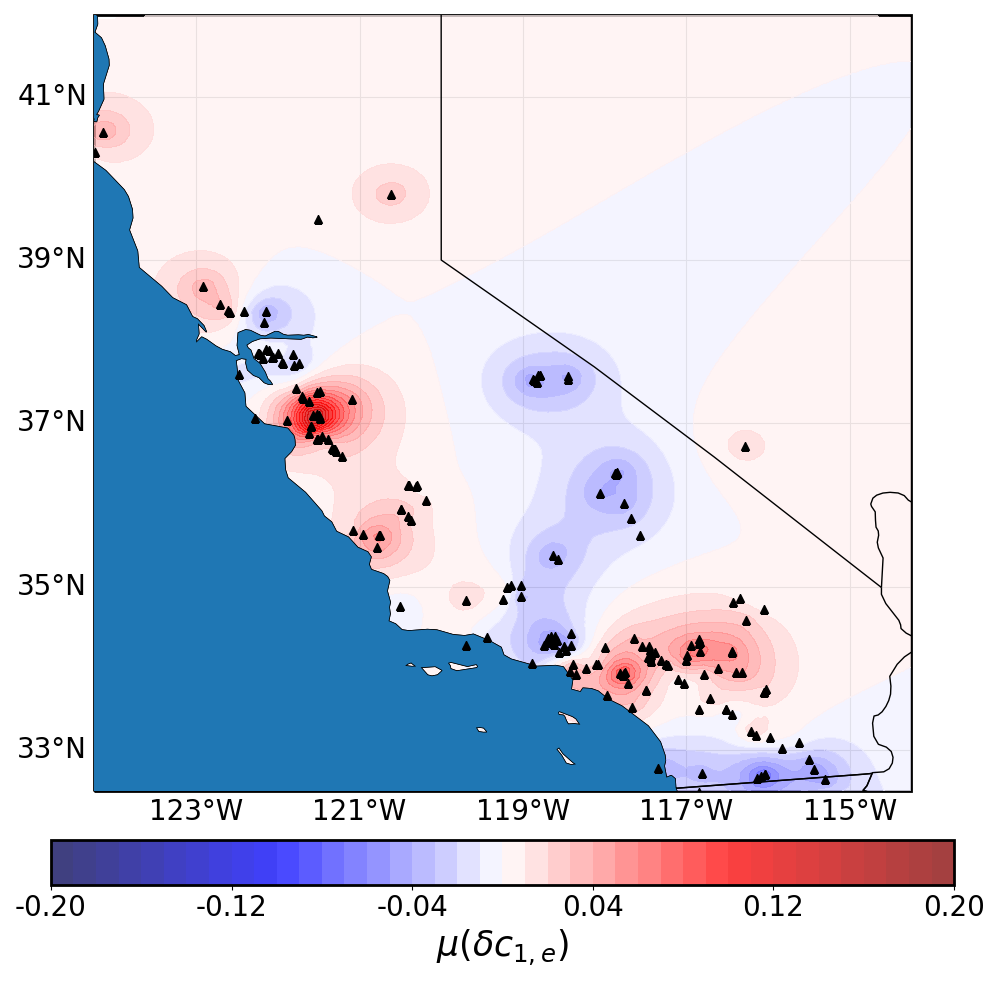}
    \end{subfigure}
    \begin{subfigure}[t]{0.4\textwidth}
        \caption{}
        \includegraphics[width = .95\textwidth]{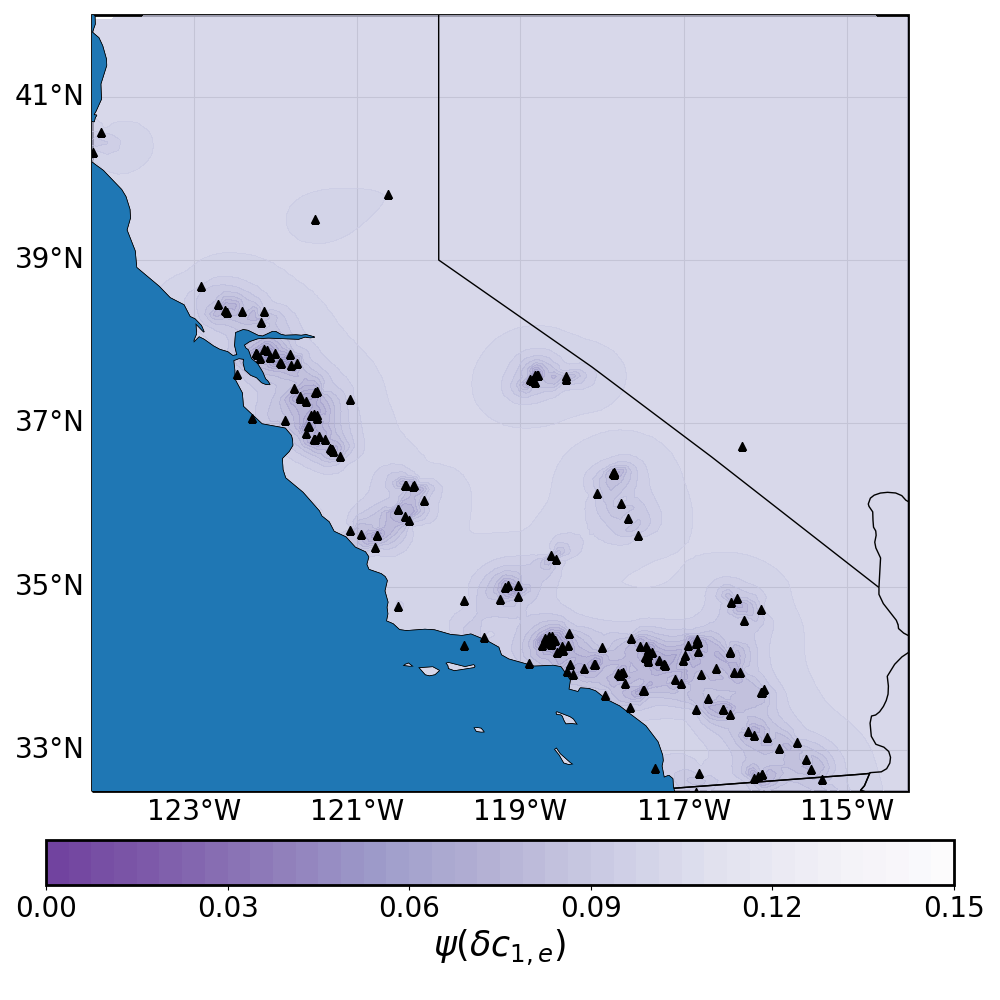}
    \end{subfigure}
    \\
    \begin{subfigure}[t]{0.4\textwidth} 
        \caption{}
        \includegraphics[width = .95\textwidth]{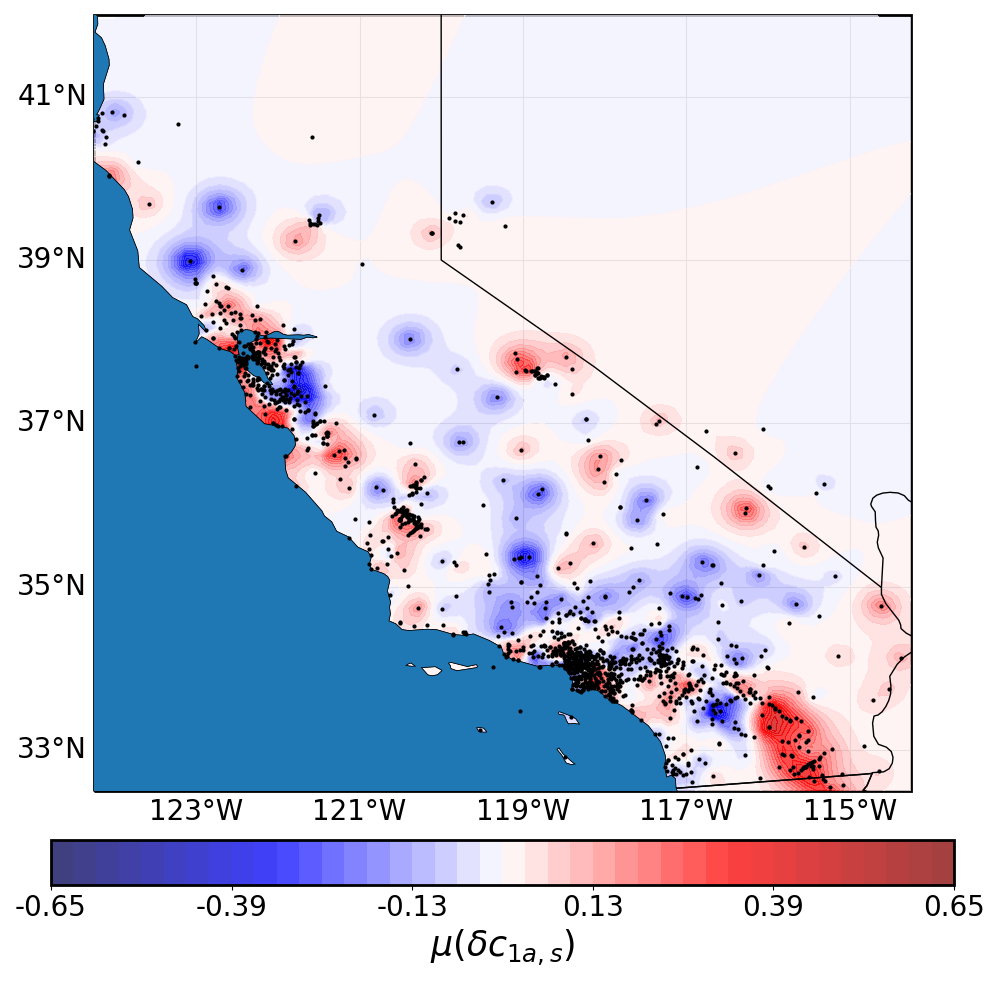}
    \end{subfigure}
    \begin{subfigure}[t]{0.4\textwidth}
        \caption{}
        \includegraphics[width = .95\textwidth]{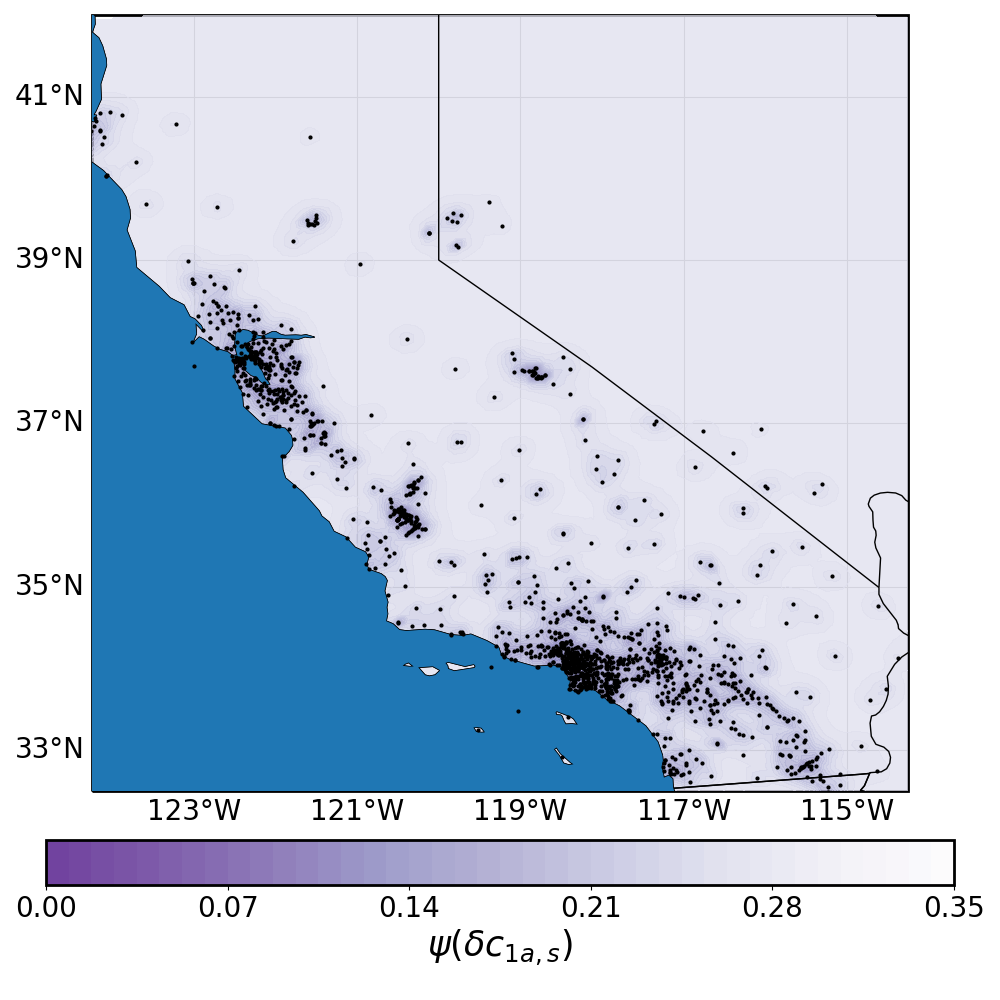}
    \end{subfigure}
    \\
    \begin{subfigure}[t]{0.4\textwidth} 
        \caption{}
        \includegraphics[width = .95\textwidth]{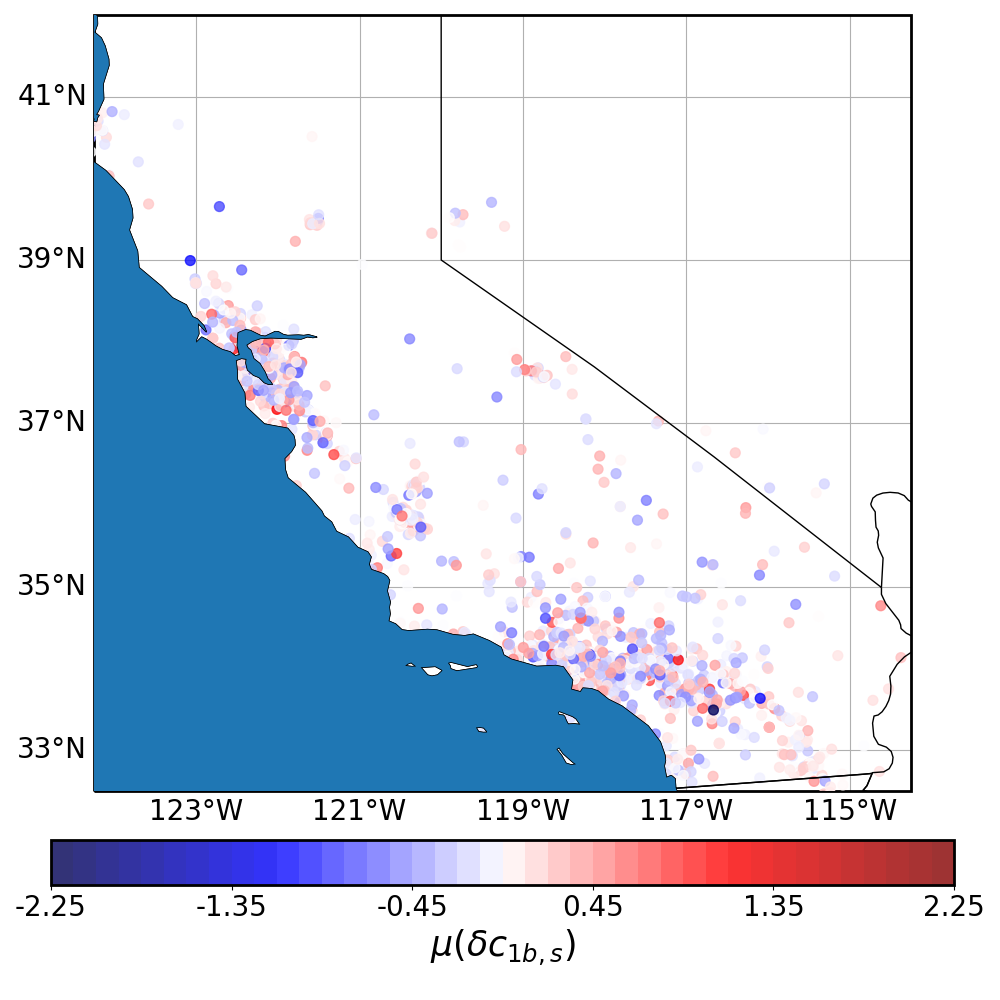}
    \end{subfigure}
    \begin{subfigure}[t]{0.4\textwidth}
        \caption{}
        \includegraphics[width = .95\textwidth]{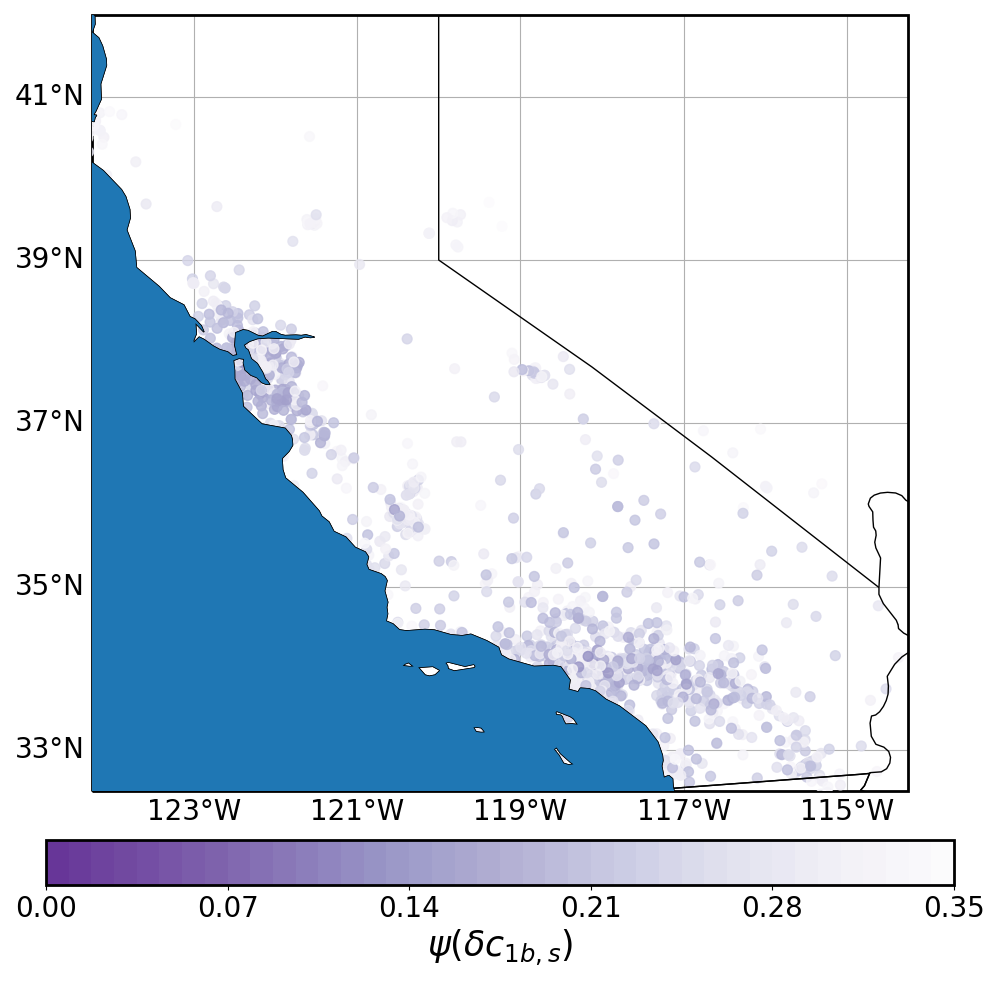}
    \end{subfigure}
    \caption{Spatial distribution of source and site constants at $f=5Hz$. Triangle markers show the location of earthquakes, dots show the location of stations.
    (a) mean estimate of $\delta c_{1,e}$, 
    (b) epistemic uncertainty of $\delta c_{1,e}$,
    (c) mean estimate of $\delta c_{1a,s}$,
    (d) epistemic uncertainty of $\delta c_{1a,s}$,
    (e) mean estimate of $\delta c_{1b,s}$, and
    (f) epistemic uncertainty of $\delta c_{1b,s}$.}
    \label{fig:spvar_coeff}
\end{figure}

Figure \ref{fig:spvar_coeff} shows the spatial distribution of the mean estimate and epistemic uncertainty of $\delta c_{1,e}$, $\delta c_{1a,s}$, and $\delta c_{1b,s}$ for $f=5Hz$.
As mentioned in Section \ref{sec:gmm_dev}, the $\delta c_{1,e}$ varies as a function of the source coordinates, whereas $\delta c_{1a,s}$ and $\delta c_{1b,s}$ as a function the site coordinates. 
In areas with past earthquakes, the mean estimate of $\delta c_{1,e}$ deviates from zero and its epistemic uncertainty is small.
$\delta c_{1,e}$ is positive if the earthquakes in a region have systematically above average source effects and negative if the source effects are below the average. 
In areas with sparse or no data, the systematic effects related to the source cannot be reliably estimated, thus, $\delta c_{1,e}$ approaches zero and its epistemic uncertainty is large.
The same behavior is observed in the spatial distribution of $\delta c_{1a,s}$: in large metropolitan areas, were most of the station are located, the $\delta c_{1a,s}$ mean estimate deviates from zero, and its epistemic uncertainty is small; in remote areas, the $\delta c_{1a,s}$ mean estimate approaches zero and, its epistemic uncertainty is large.
$\delta c_{1b,s}$ is only plotted at the station locations as it has a zero correlation length, meaning that as we move away from a station it will directly go to zero.
The mean estimates of $\delta c_{1b,s}$ do not exhibit any spatial correlation (i.e. there are no regions where $\delta c_{1b,s}$ are systematically positive or negative) meaning that spatially correlated component of the site effects was properly captured by $\delta c_{1a,s}$

\begin{figure}
    \centering
    \begin{subfigure}[t]{0.48\textwidth} 
        \caption{} \label{fig:cA_cell_mu}
        \includegraphics[width = .95\textwidth]{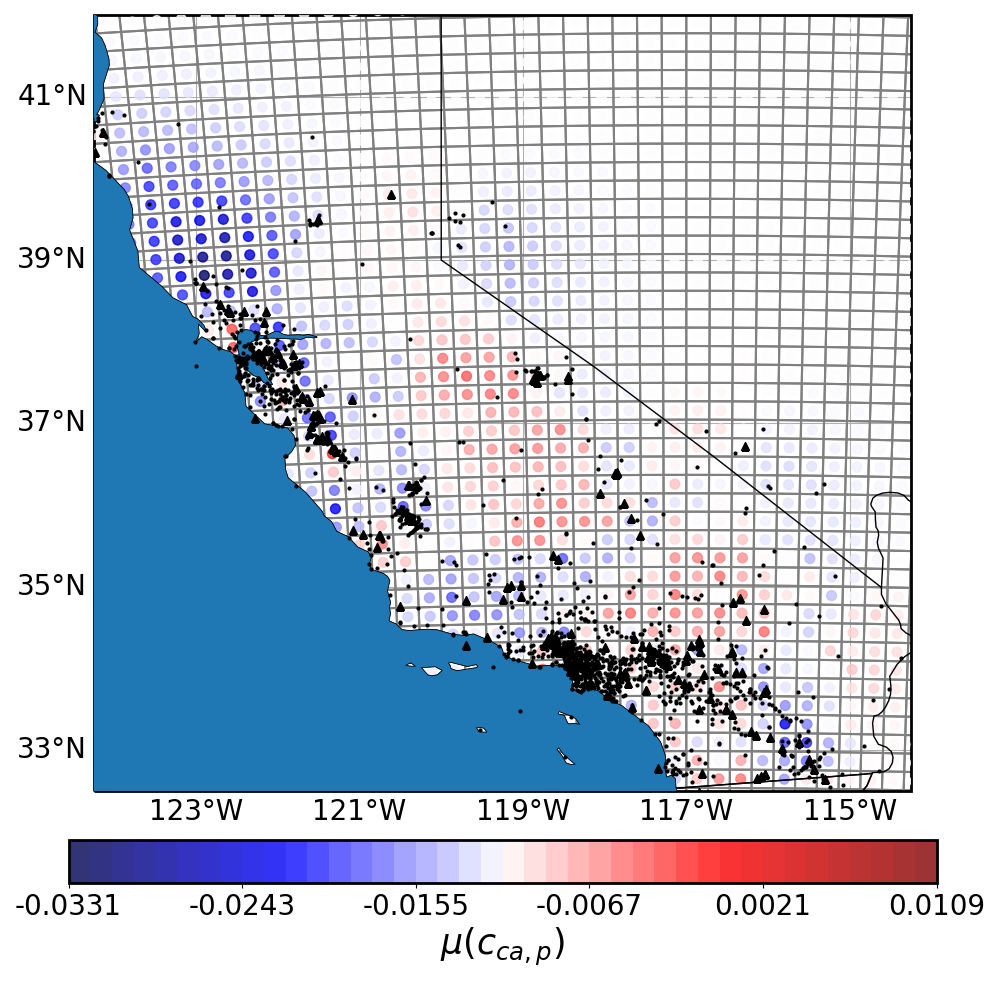}
    \end{subfigure}
    \begin{subfigure}[t]{0.48\textwidth}
        \caption{}
        \includegraphics[width = .95\textwidth]{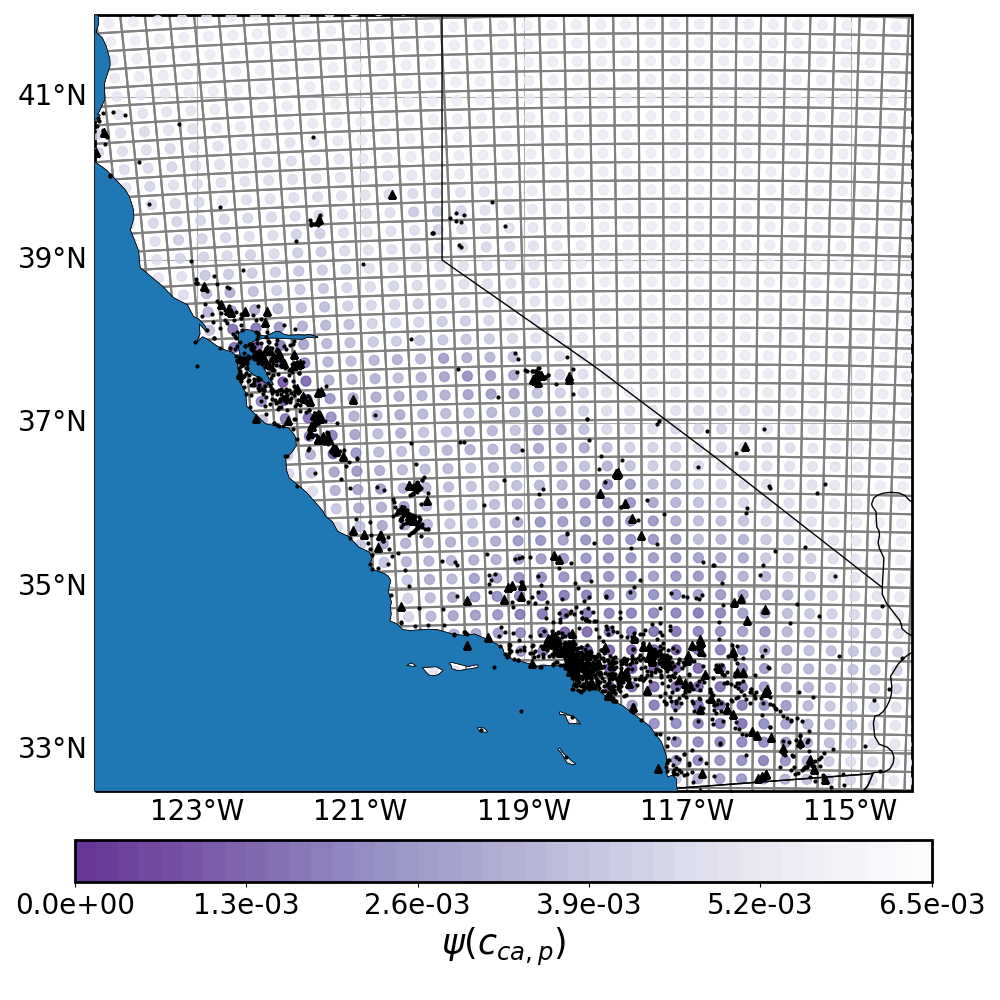}
    \end{subfigure}
    \caption{Spatial distribution of cell specific anelastic attenuation at $f=5Hz$. 
    (a) mean estimate of $c_{ca,p}$, and
    (b) epistemic uncertainty of $c_{ca,p}$. }
    \label{fig:cA_cell}
\end{figure}

Figure \ref{fig:cA_cell} illustrates the spatial distribution of the cell-specific anelastic attenuation. 
The mean of $c_{ca,p}$ deviates from $c_{7~BA18}$ in cells that are crossed by many paths, whereas it stays close to $c_{7~BA18}$ in cells crossed by few or zero paths. 
In addition, cells that are crossed by few paths have large epistemic uncertainty in $c_{ca,p}$.
Overall, the epistemic uncertainty is low in Bay Area and Los Angeles and high in the northern part of California and the state of Nevada.
The main features that stand out in Figure \ref{fig:cA_cell_mu} are the higher than average anelastic attenuation north of the San Francisco Bay Area and east of San Diego, and the less than average anelastic attenuation in the Central Valley and east of Los Angeles.

\begin{figure}
    \centering
    \begin{subfigure}[t]{0.48\textwidth} 
        \caption{} \label{fig:Q_Phillips}
        \includegraphics[width = .95\textwidth]{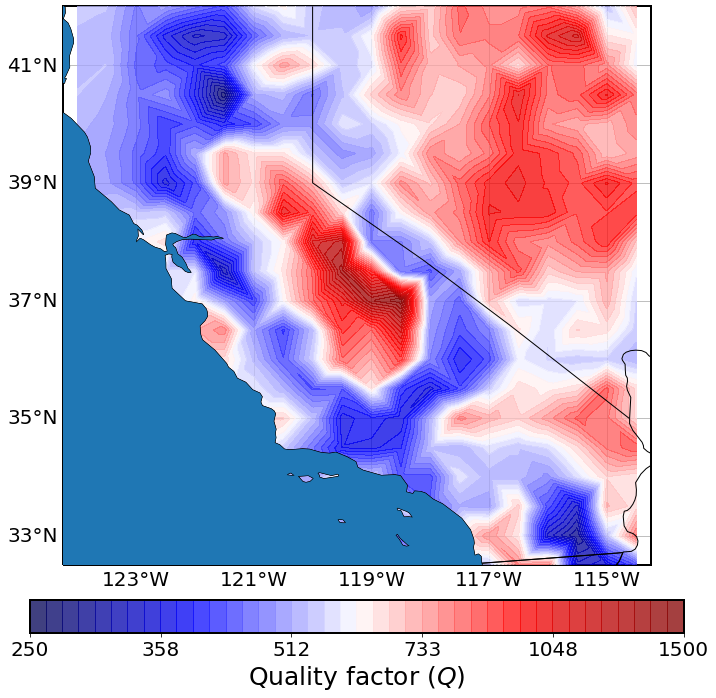}
    \end{subfigure}
    \begin{subfigure}[t]{0.48\textwidth}
        \caption{} \label{fig:Q_Eberhart}
        \includegraphics[width = .95\textwidth]{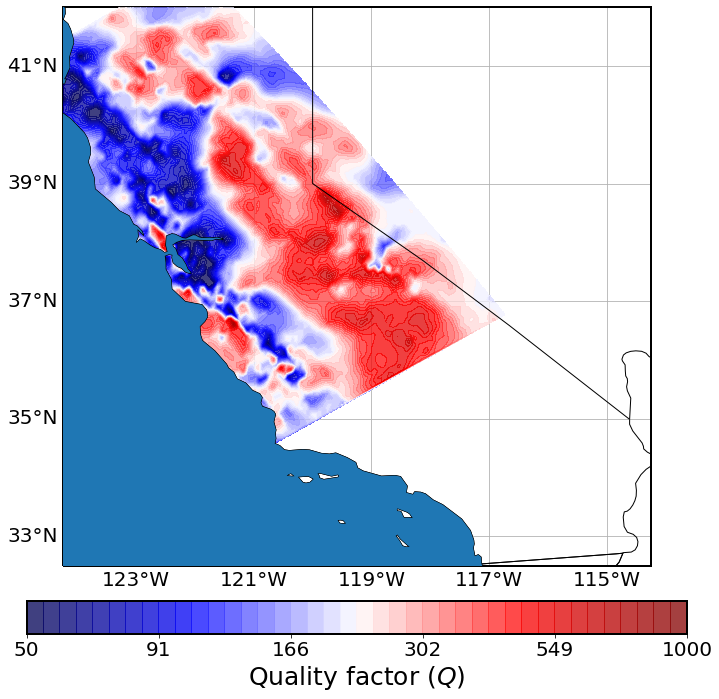}
    \end{subfigure}
    \caption{Seismic attenuation models for California from seismic inversions. (a) \cite{Phillips2014}  attenuation model for frequencies between $6$ and $12~Hz$, and (b) \cite{Eberhart-Phillips2016} S-wave attenuation model for northern California at depth of $4 km$.}
    \label{fig:Q_map}
\end{figure}

These findings are consistent with published attenuation models shown in Figure \ref{fig:Q_map}; Figure \ref{fig:Q_Phillips} corresponds to the \cite{Eberhart-Phillips2016} $Q$ model for frequencies $6$ to $12~Hz$, and \ref{fig:Q_Phillips} corresponds to the \cite{Phillips2014} $Q$ model for the S-waves at $4km$ depth.
The quality factor, $Q$, is inversely proportional to the anelastic attenuation: high $Q$ means low anelastic attenuation, and vice versa. 
Both models show small $Q$ values north of Bay Area, and large $Q$ values in Central Valley; additionally, \cite{Eberhart-Phillips2016}, which covers the entire state of California and Nevada, shows small values of $Q$ east of San Diego and large values of $Q$ east of Los Angeles.
The mean value of $c_{ca,p}$ and the $Q$ model of \cite{Eberhart-Phillips2016} differ in Nevada because there are no paths that cover that region.
The large epistemic uncertainty of $c_{ca,p}$ in Nevada means that the cell-specific anelastic attenuation cannot be estimated in that region with the current data set.
This comparison shows that the cell specific anelastic attenuation captures an underlying physical behavior. 

\begin{figure}
    \centering
    \begin{subfigure}[t]{0.4\textwidth} 
        \caption{}
        \includegraphics[width = .95\textwidth]{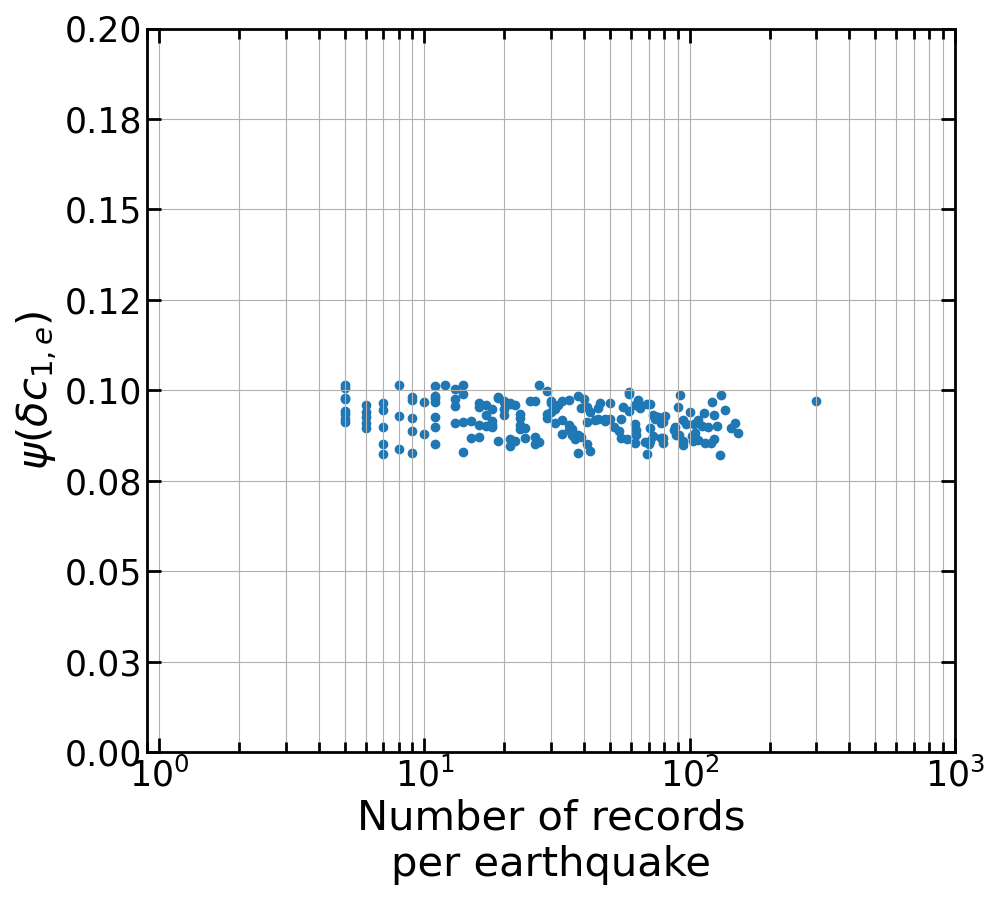}
    \end{subfigure}
    \begin{subfigure}[t]{0.4\textwidth}
        \caption{}
        \includegraphics[width = .95\textwidth]{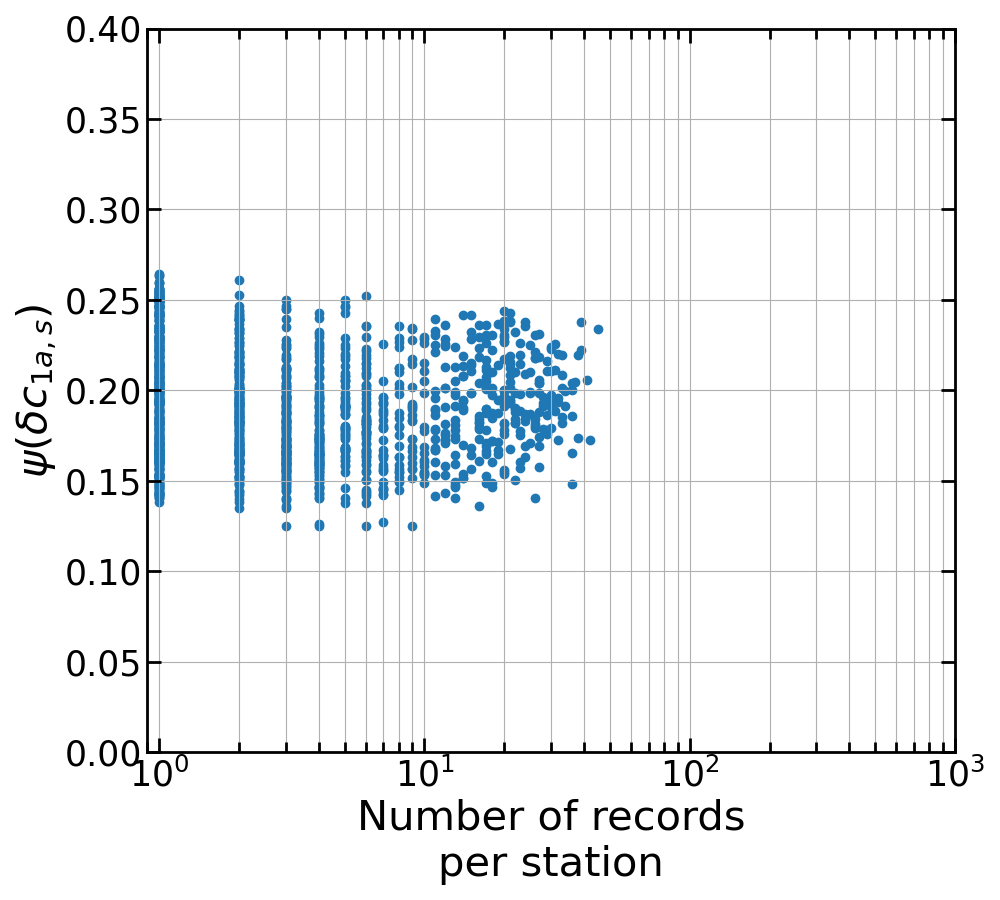}
    \end{subfigure}
    \\
    \begin{subfigure}[t]{0.4\textwidth} 
        \caption{}
        \includegraphics[width = .95\textwidth]{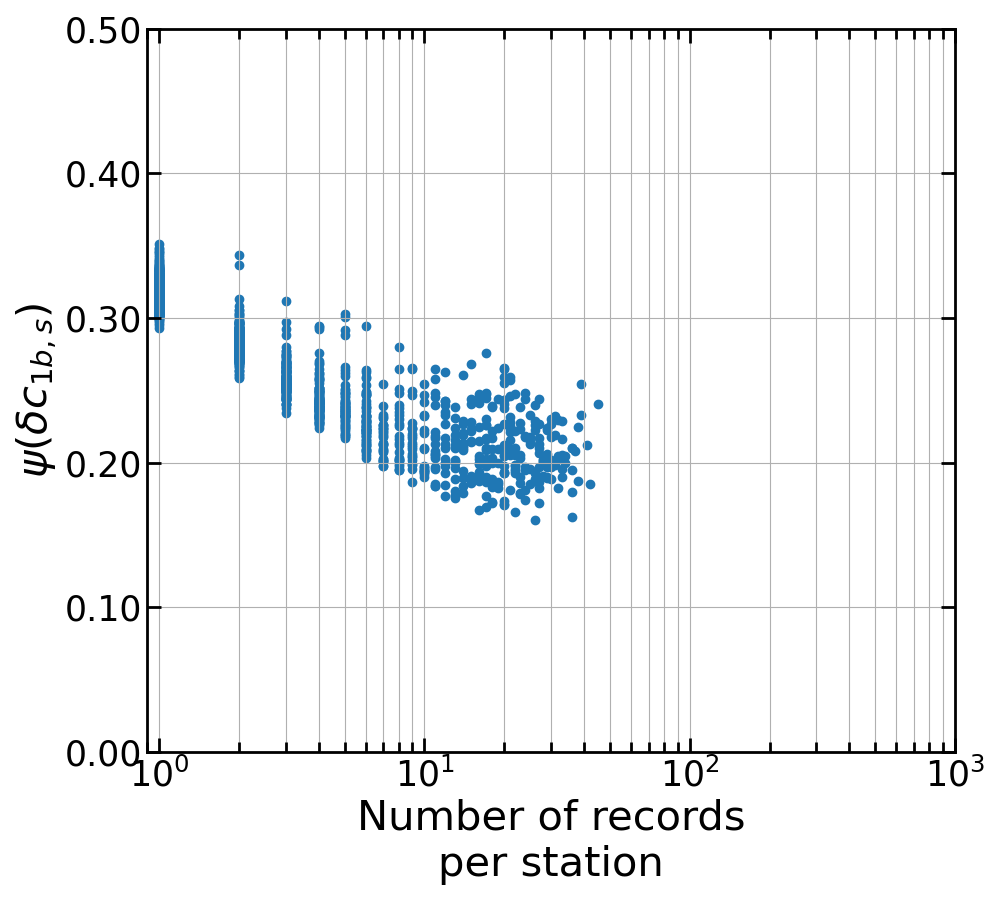}
    \end{subfigure}
    \begin{subfigure}[t]{0.4\textwidth}
        \caption{}
        \includegraphics[width = .95\textwidth]{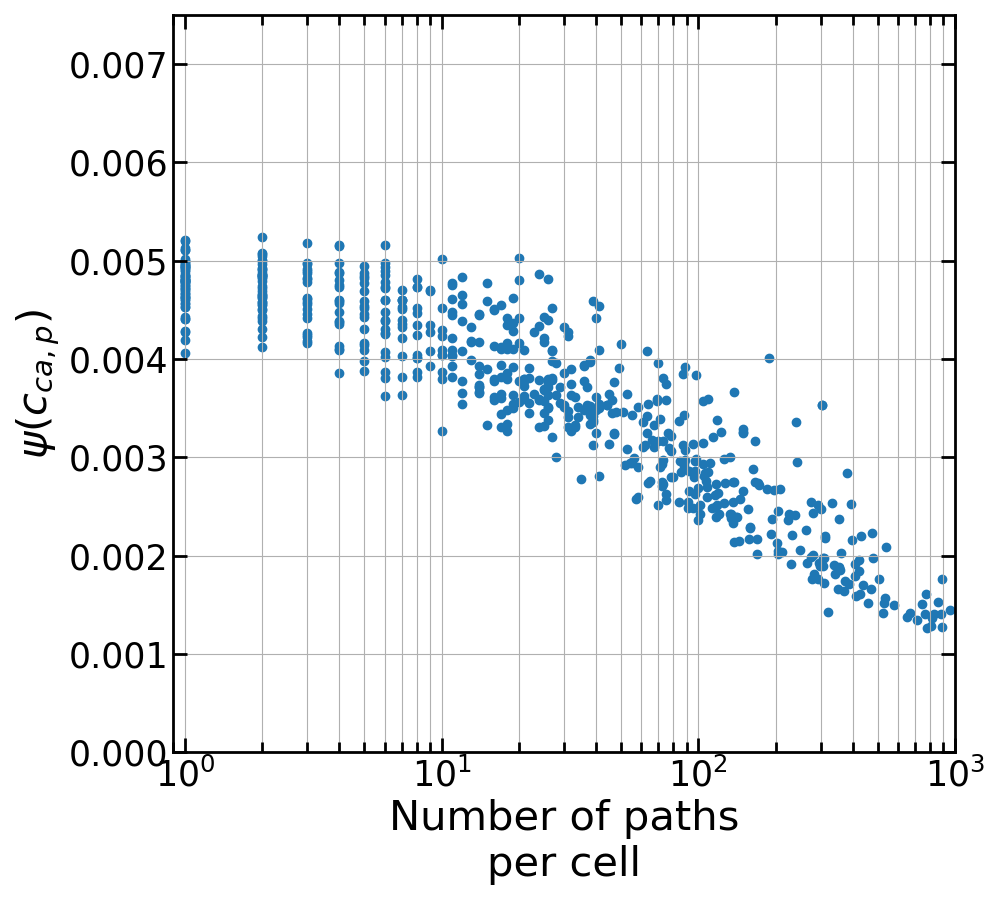}
    \end{subfigure}
    \caption{Standard deviations of posterior distributions of non-ergodic terms; 
    (a) $\delta c_{1,e}$, 
    (b) $\delta c_{1a,s}$
    (c) $\delta c_{1b,s}$, and
    (d) $c_{ca,p}$.}
    \label{fig:unc_nerg_terms}
\end{figure}

Figure \ref{fig:unc_nerg_terms} shows the epistemic uncertainty of the non-ergodic terms as a function of the number of records for $\delta c_{1,e}$, $\delta c_{1a,s}$ and $\delta c_{1b,s}$, and as a function of the number of paths for $c_{ca,p}$. 
The epistemic uncertainty of $\delta c_{1,e}$ and $\delta c_{1a,s}$ is not sensitive to the number of records, whereas the epistemic uncertainty of $\delta c_{1b,s}$ and $c_{ca,p}$ decreases as the number of records and number of paths increases. 
This occurs because $\delta c_{1b,s}$ is spatially uncorrelated, and $c_{ca,p}$ has a spatially uncorrelated component; $\delta c_{1b,s}$ can be estimated more accurately as the number of ground motions recorded at a station increases, and $c_{ca,p}$ can be estimated more accurately as the number of paths crossing a cell increases. 
$\delta c_{1,e}$ and $\delta c_{1a,s}$ are spatially correlated and so the location of an event or a station is also important. 
That is, $\delta c_{1a,s}$ can have less epistemic uncertainty near a group of stations with few records at each station than near a remote station with a large number of records, if collectively, the group of stations has more data to constrain $\delta c_{1a,s}$.
The same holds true for $\delta c_{1,e}$ regarding the spatial distribution of events.

\subsection{Non-ergodic residuals}

\begin{figure}
    \centering
    \begin{subfigure}[t]{0.40\textwidth} 
        \caption{} 
        \includegraphics[width = .95\textwidth]{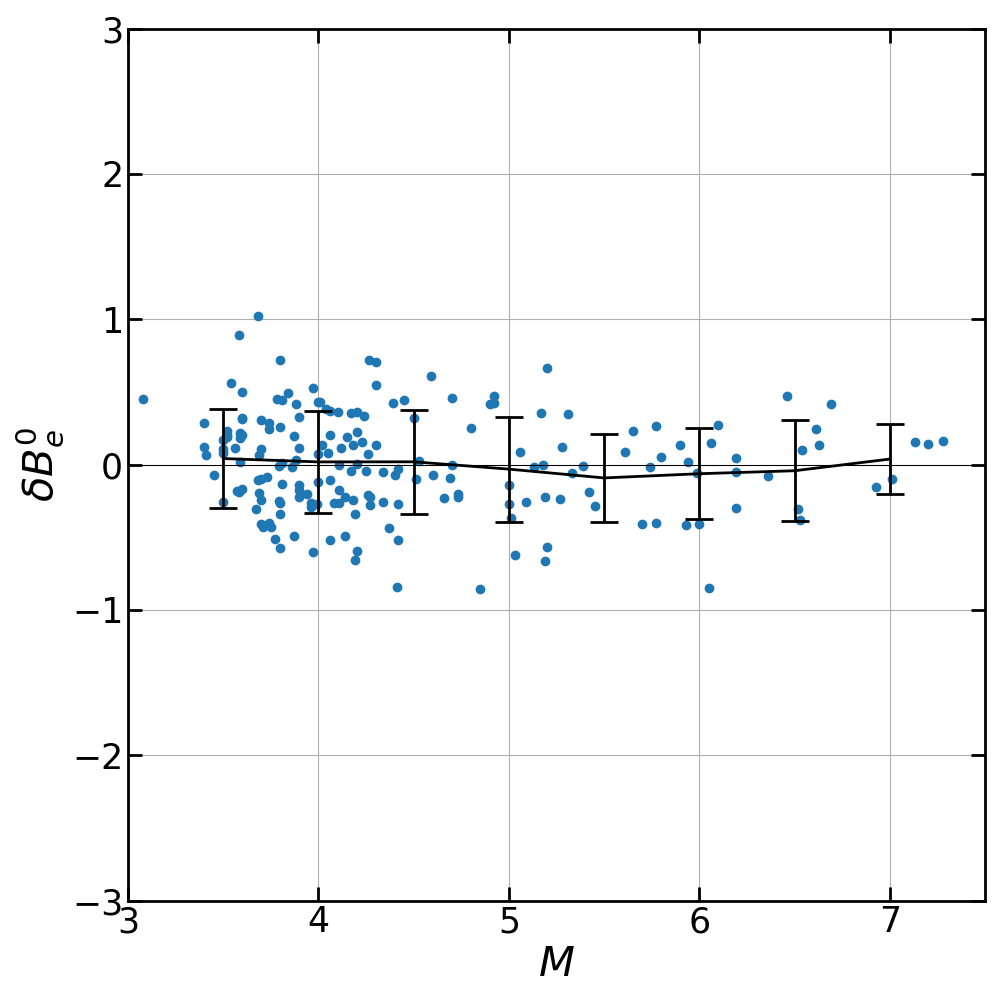}
    \end{subfigure}
    \begin{subfigure}[t]{0.40\textwidth}
        \caption{} 
        \includegraphics[width = .95\textwidth]{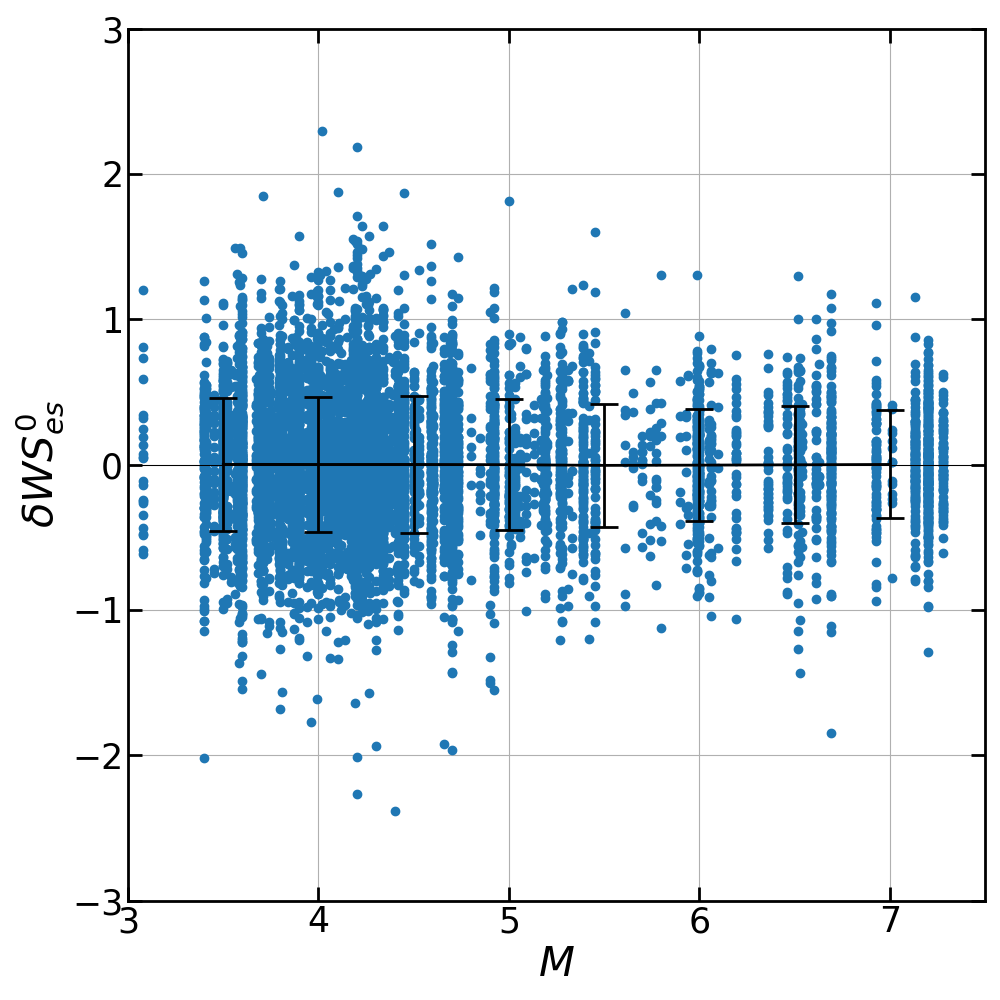}
    \end{subfigure}
    \\
    \begin{subfigure}[t]{0.40\textwidth} 
        \caption{} 
        \includegraphics[width = .95\textwidth]{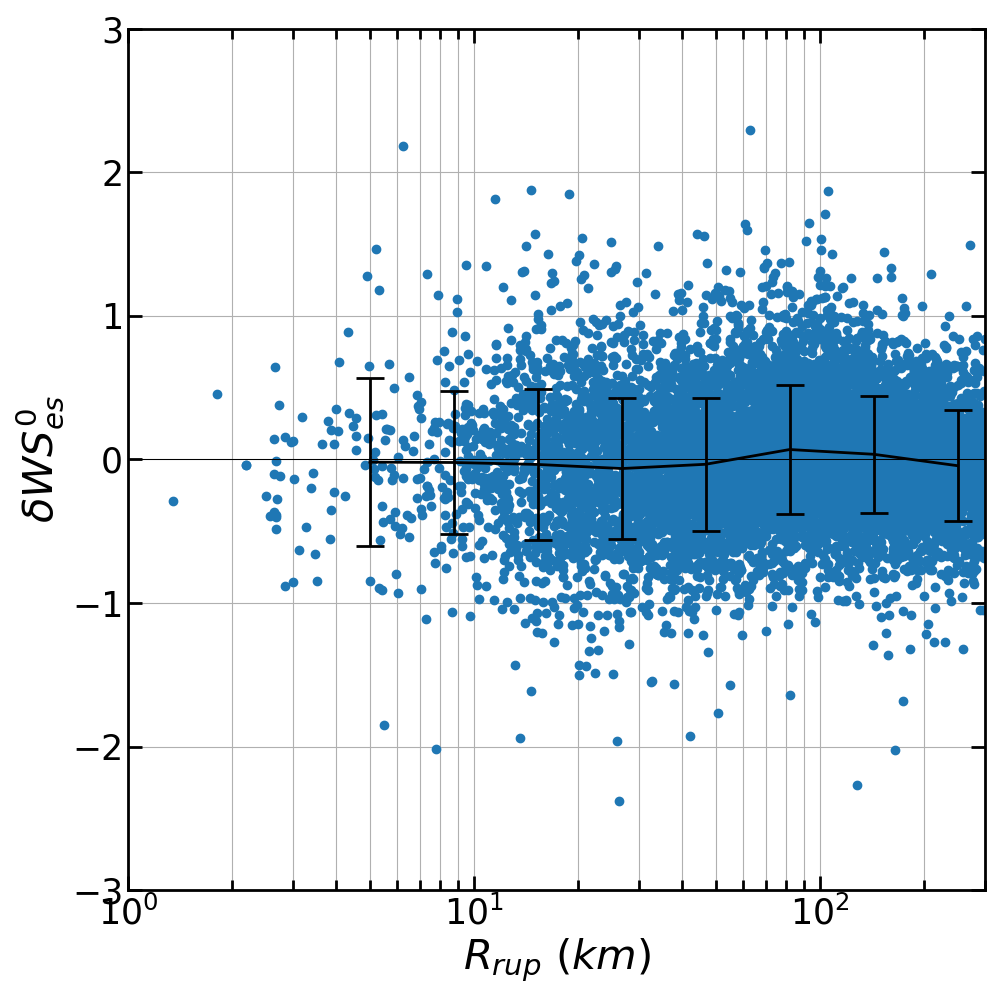}
    \end{subfigure}
    \begin{subfigure}[t]{0.40\textwidth}
        \caption{}
        \includegraphics[width = .95\textwidth]{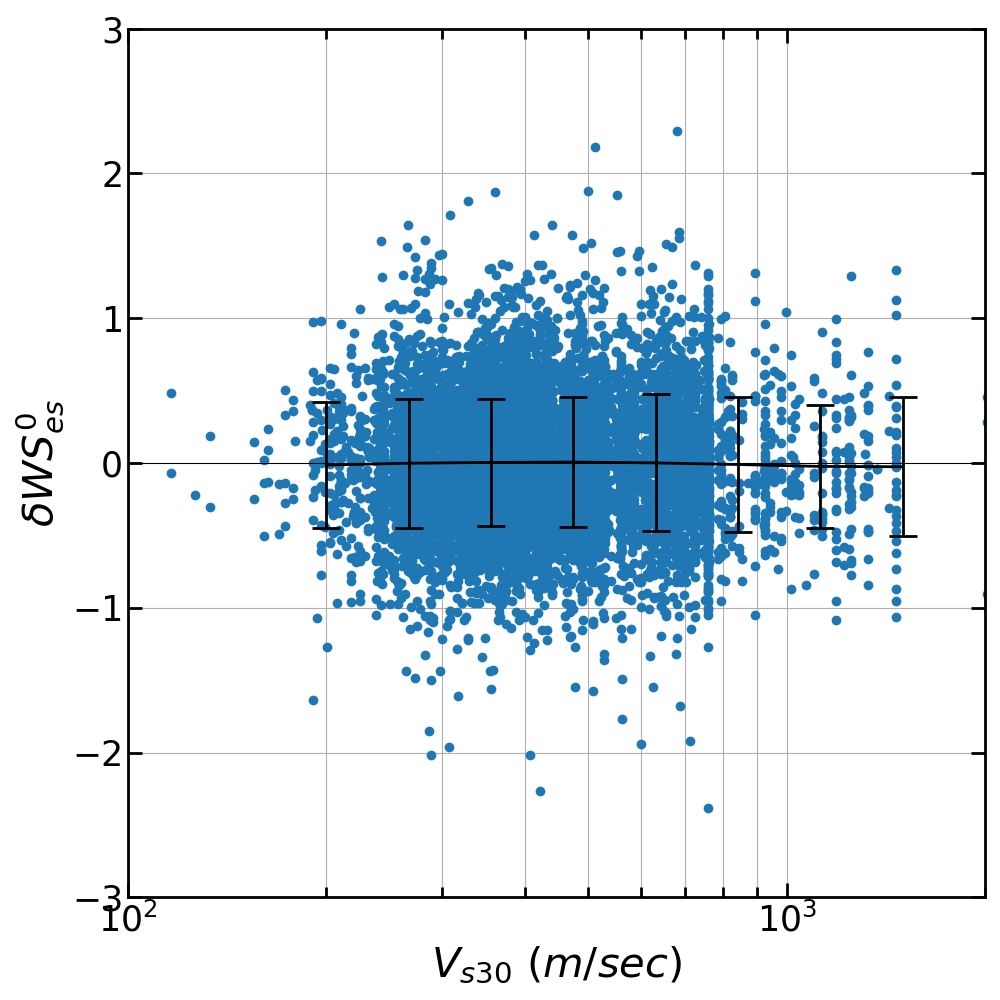}
    \end{subfigure}
    \caption{Non-ergodic within-event and between-event residuals at $f=5Hz$. (a) $\delta B^0_e$ versus magnitude, (b) $\delta WS^0_{es}$ versus magnitude, (c) $\delta WS^0_{es}$ versus rupture distance, and (d) $\delta WS^0_{es}$ versus $V_{S30}$.}
    \label{fig:res}
\end{figure}

The residuals of the non-ergodic model at $f=5Hz$ are presented in Figure \ref{fig:res}: the dots represent the residuals, the solid line corresponds to the moving average, and the error bars correspond to the standard deviation.
$\delta B^0_e$ shows no trend and an approximately constant standard deviation with magnitude; $\delta WS^0_{es}$ also shows no trend, but the standard deviation standard deviation reduces with magnitude.
Additionally, $\delta WS^0_{es}$ shows no trend and a constant standard deviation with $R_{rup}$ and $V_{S30}$

\subsection{Standard deviation} \label{sec:std}

In the model development, for simplicity, the aleatory standard deviations, $\tau_0$ and $\phi_0$, were modeled as magnitude independent. 
Any magnitude dependence of $\tau_0$ and $\phi_0$ was determined in post processing based on the non-ergodic residuals. 
$\tau_0$ was modeled as constant, as, for the most part, the $\delta B^0_{e}$ residuals did not exhibit any reduction in standard deviation with magnitude (Figure \ref{fig:res}).
Because the number of events greater than $M~6.5$ is small, the model for $\tau_0$ did not follow the reduction of the empirical standard deviation at large magnitudes, but instead it followed the standard deviation of the small events. 
$\phi_0$ was modeled as a piece-wise linear function (equation \eqref{eq:phi_model}), as $\delta W^0_{es}$ residuals exhibit some reduction in the standard deviation with increasing magnitude: $\phi_{0M_1}$ is the within-event standard deviation for magnitudes less than $5$, and $\phi_{0M_2}$ is the within-event standard deviation for magnitudes greater than $6.5$.

\begin{figure}
    \centering
    \includegraphics[width = .40\textwidth]{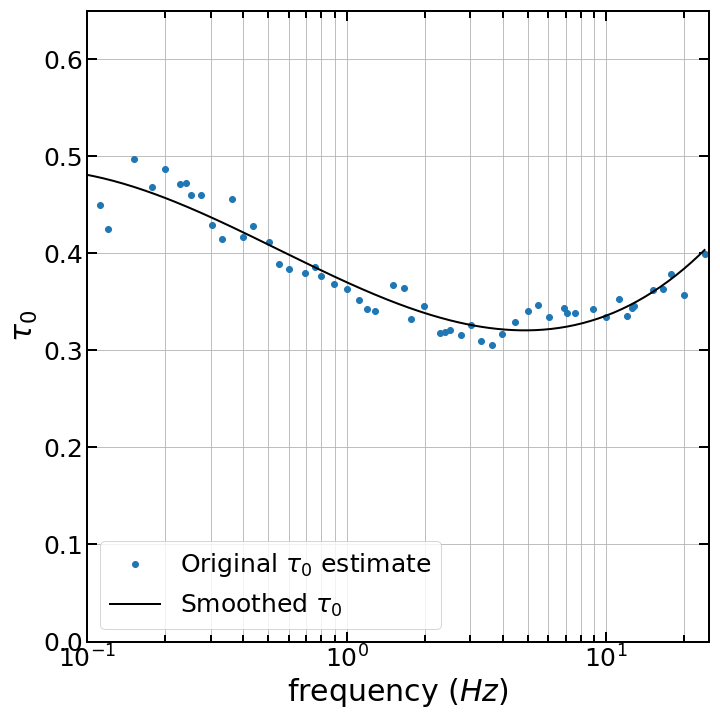}
    \caption{Between-event standard deviation, $\tau_0$, versus frequency; circular markers correspond to the estimated $\tau_0$ at every frequency, solid line corresponds to smoothed $\tau_0$. }
    \label{fig:tau_model}
\end{figure}

\begin{figure}
    \centering
    \includegraphics[width = .40\textwidth]{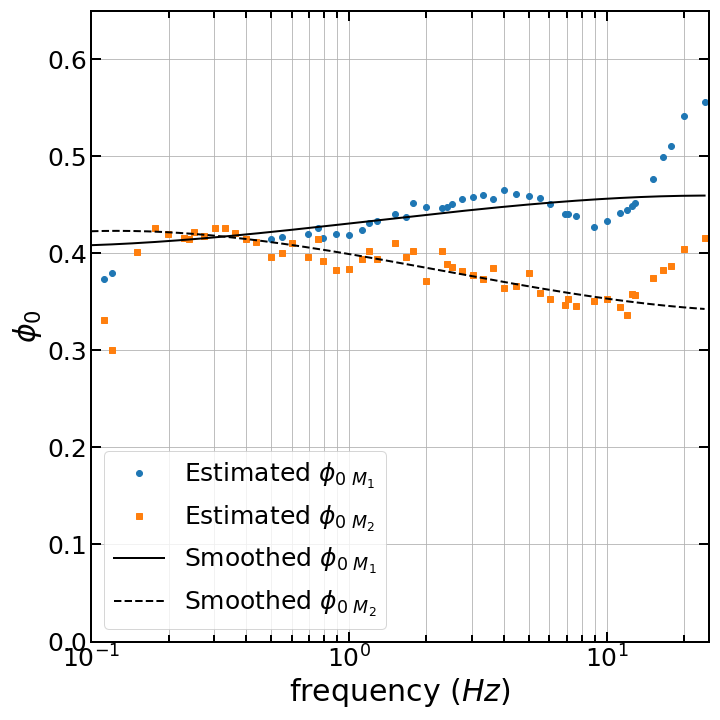}
    \caption{Within-event standard deviation, $\phi_0$, versus frequency; 
    circular markers correspond to the estimated $\phi_0$ at small magnitudes, square markers correspond to the estimated  $\phi_0$ at large magnitudes, solid line corresponds to smoothed $\phi_0$ for small magnitudes, dashed line corresponds to the smoothed $\phi_0$ for large magnitudes.}
    \label{fig:phi_model}
\end{figure}

\begin{figure}
    \centering
    \begin{subfigure}[t]{0.40\textwidth} 
        \caption{} 
        \includegraphics[width = .95\textwidth]{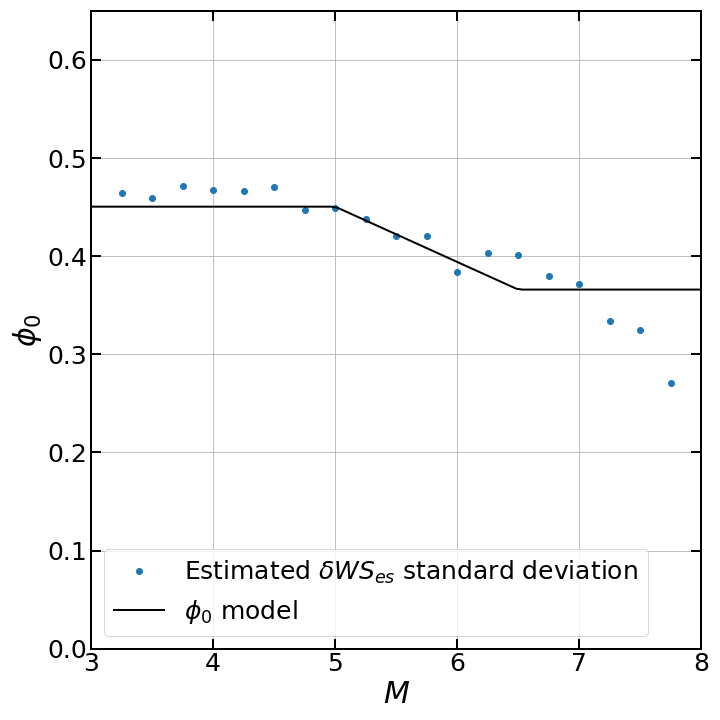}
    \end{subfigure}
    \begin{subfigure}[t]{0.40\textwidth}
        \caption{} 
        \includegraphics[width = .95\textwidth]{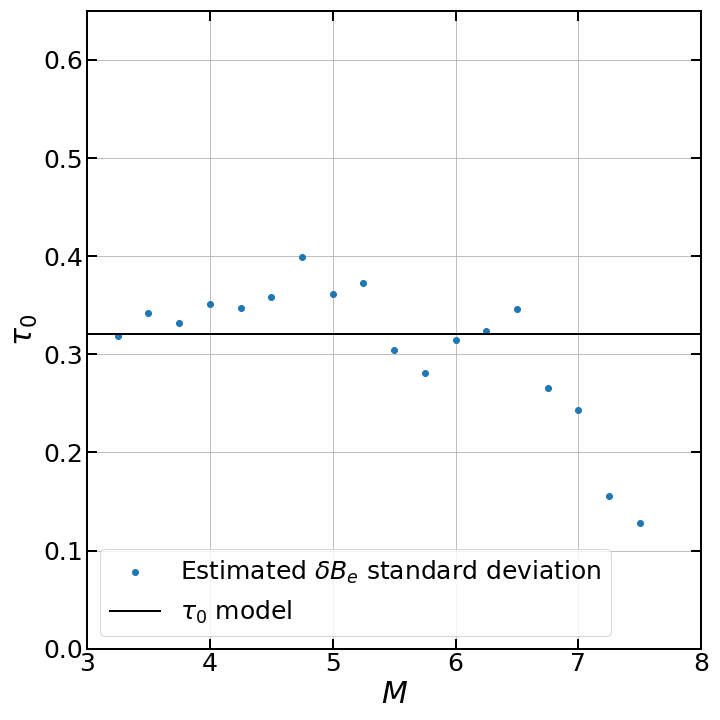}
    \end{subfigure}
    \caption{Magnitude scaling of $\tau_0$ and $\phi_0$ for $f=5Hz$; circular markers denote the standard deviations of the binned residuals, and solid lines correspond to the standard deviation models.}
    \label{fig:alet_vs_mag}
\end{figure}

The aleatory parameters ($\tau_0$, $\phi_{0M_1}$, and $\phi_{0M_2}$) were smoothed in order to ensure that the resulting $EAS$ will have a reasonable shape (Figures \ref{fig:tau_model} and \ref{fig:phi_model}).
The smoothing was performed by fitting the aleatory parameters with a fourth order polynomial.
The value of $\tau_0$ decreases from small to intermediate frequencies and increases again after $f=3Hz$, which is consistent with the behaviour of BA18 and other $PSA$ GMM, such as \cite{Abrahamson2014}.
The magnitude dependence of $\phi_0$ is more pronounced at high frequencies. 
The higher $\phi_0$ of small events at high frequencies is believed to be due to an increased effect of the radiation patterns. 
At large events, the effect of radiation patterns is smaller as seismic rays originate from more locations along the fault, which increases the range of azimuthal angles, and leads to destructive interference of the radiation patterns resulting in less ground-motion variability.
Figure \ref{fig:alet_vs_mag} compares the magnitude relationships for $\tau_0$ and $\phi_0$ with the empirical standard deviations at $f=5Hz$.
Overall, there is a good fit between the $\tau_0$ and $\phi_0$ relationships and the standard deviations of the binned residuals. 
The differences at large magnitudes should be reevaluated with a dataset which includes a grater number of large magnitude events.

\begin{equation} \label{eq:phi_model}
    \phi_0 = \left\{ \begin{array}{cll}
                \phi_{0M_1}                                             & for & M < 5  \\
                \phi_{0M_1} + (\phi_{0M_2}-\phi_{0M_2}) (M-5)/(6.5-5)   & for & 5 < M < 6.5  \\
                \phi_{0M_2}                                             & for & M > 6.5  \\
    \end{array} 
    \right. 
\end{equation}

\begin{figure}
    \centering
    \includegraphics[width = .40\textwidth]{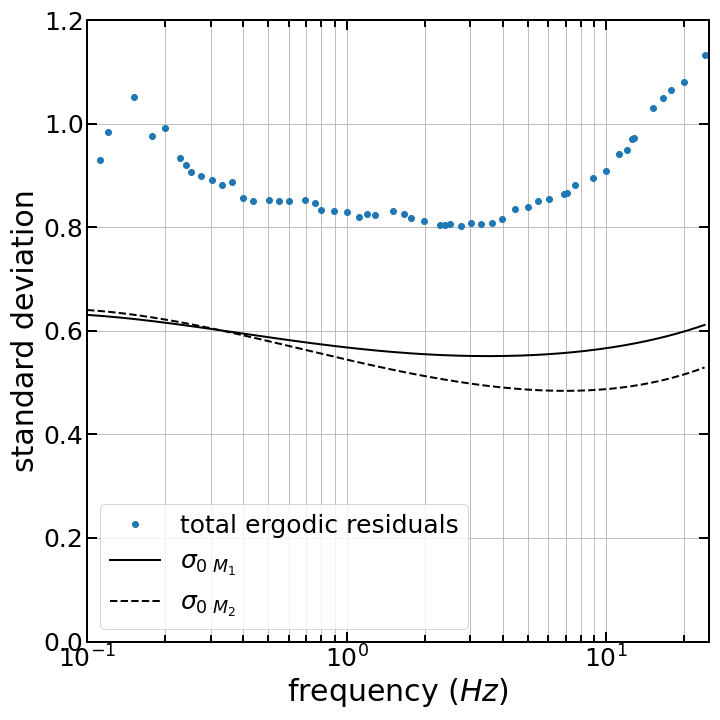}
    \caption{Comparison of total non-ergodic standard deviation with standard deviations of the total ergodic redisulas}
    \label{fig:aleat_tot}
\end{figure}

As a comparison with ergodic aleatory variability, figure \ref{fig:aleat_tot} shows the total aleatory variability for the non-ergodic GMM for the small and large magnitudes events and the standard deviation of the ergodic residuals used for the derivation of this model. 
Based on this range, the total aleatory standard deviation of the non-ergodic GMM ranges from $0.50$ to $0.65$ which is about a $40$ to $30\%$ reduction from the standard deviation of the ergodic GMM.

\subsection{Inter-frequency correlation} \label{sec:ifreq_corr}

\begin{figure}
    \centering
    \begin{subfigure}[t]{0.45\textwidth} 
        \caption{} \label{fig:ifreq_corr_c1a}
        \includegraphics[width = .95\textwidth]{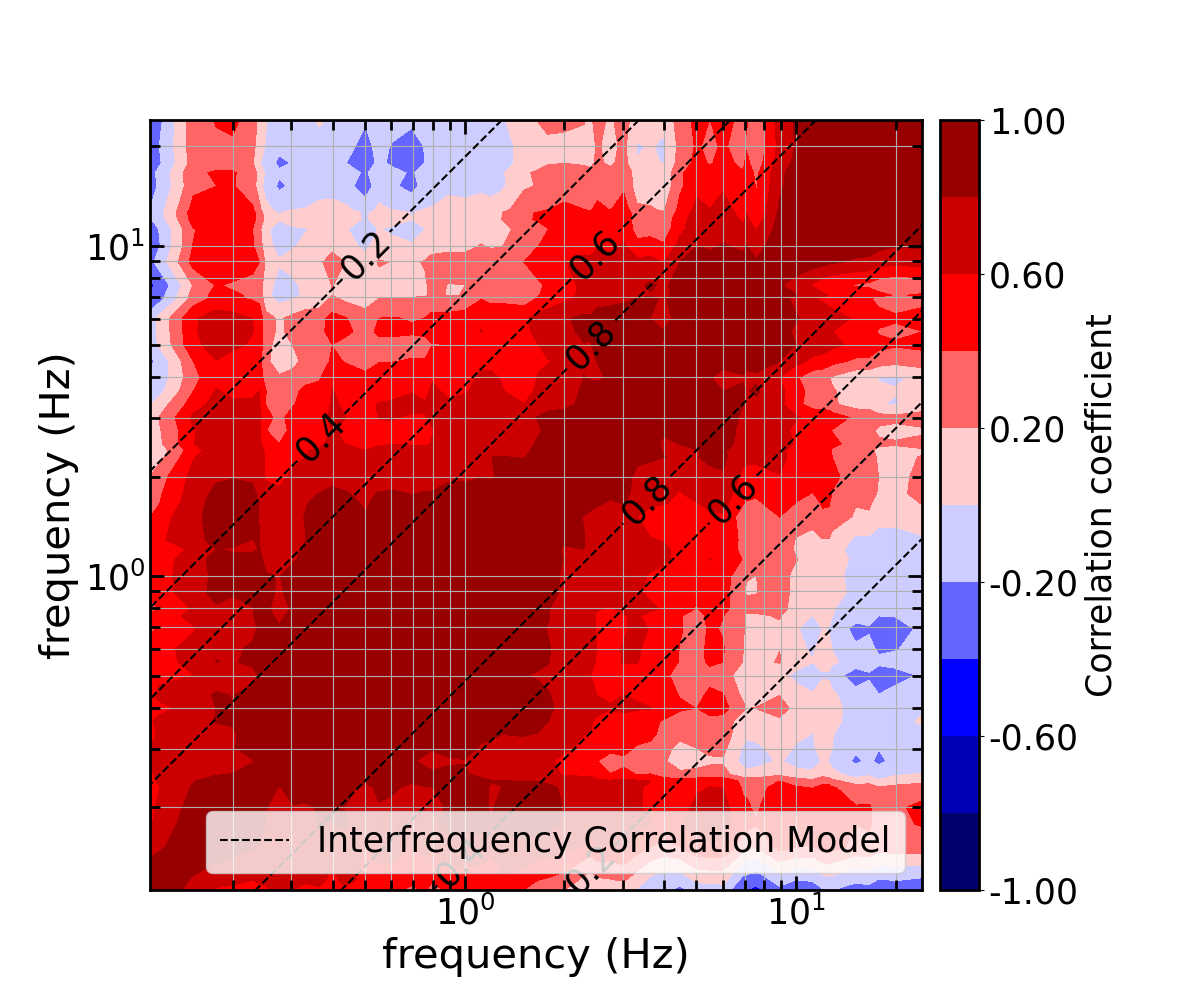}
    \end{subfigure}
    \begin{subfigure}[t]{0.45\textwidth}
        \caption{} \label{fig:ifreq_corr_c1b}
        \includegraphics[width = .95\textwidth]{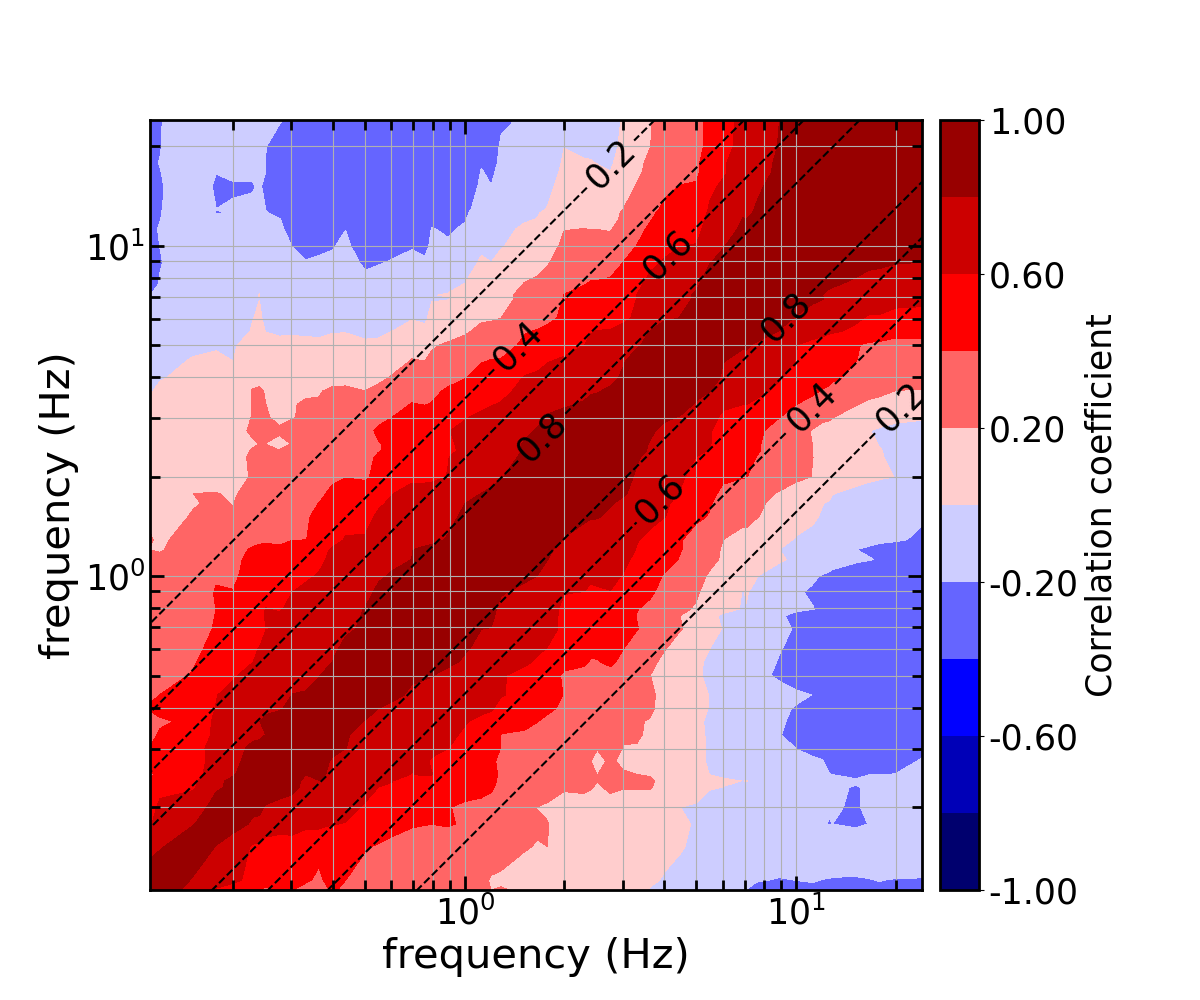}
    \end{subfigure}
    \\
    \begin{subfigure}[t]{0.45\textwidth} 
        \caption{} \label{fig:ifreq_corr_dS2S}
        \includegraphics[width = .95\textwidth]{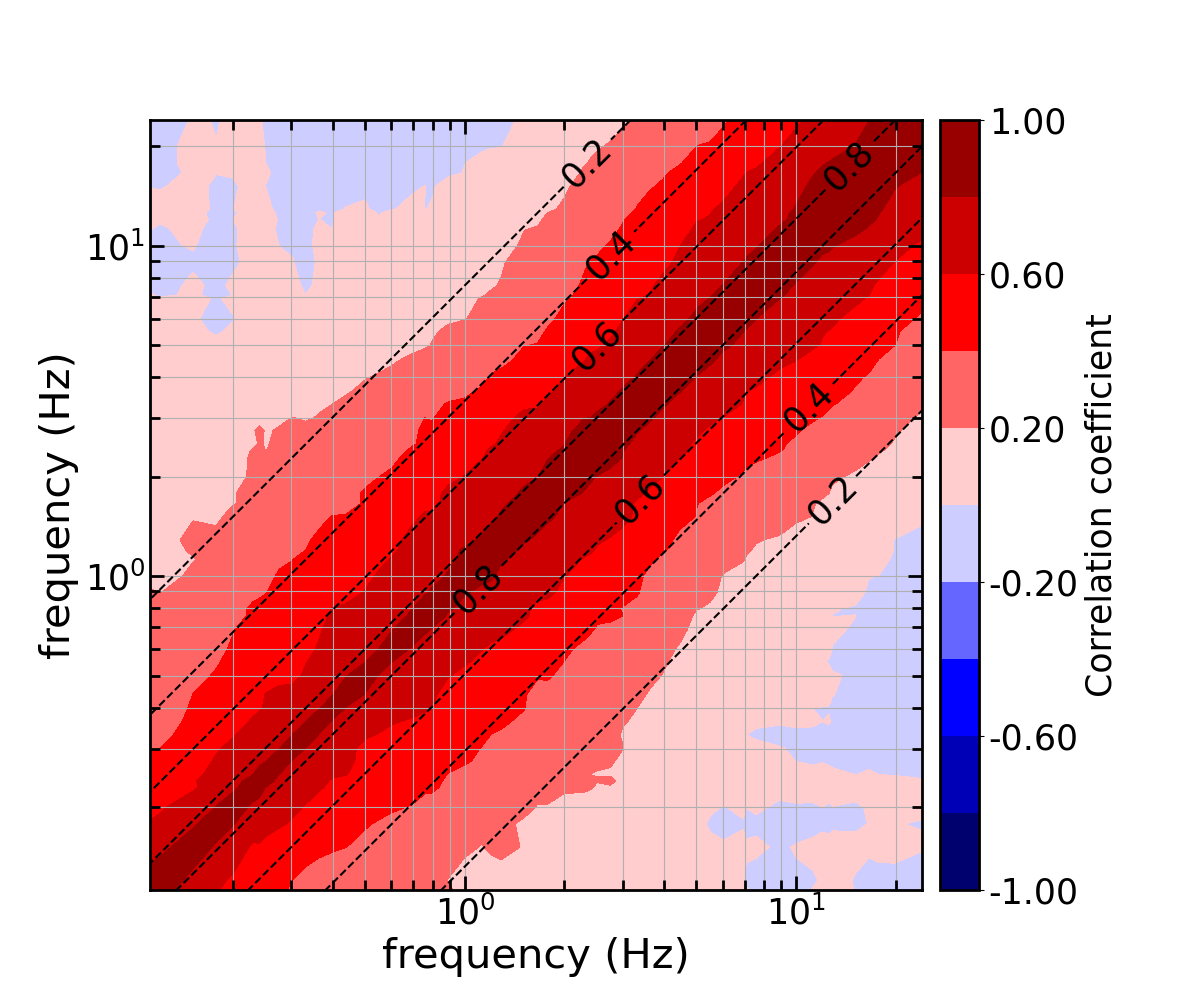}
    \end{subfigure}
    \begin{subfigure}[t]{0.45\textwidth}
        \caption{}
        \includegraphics[width = .95\textwidth]{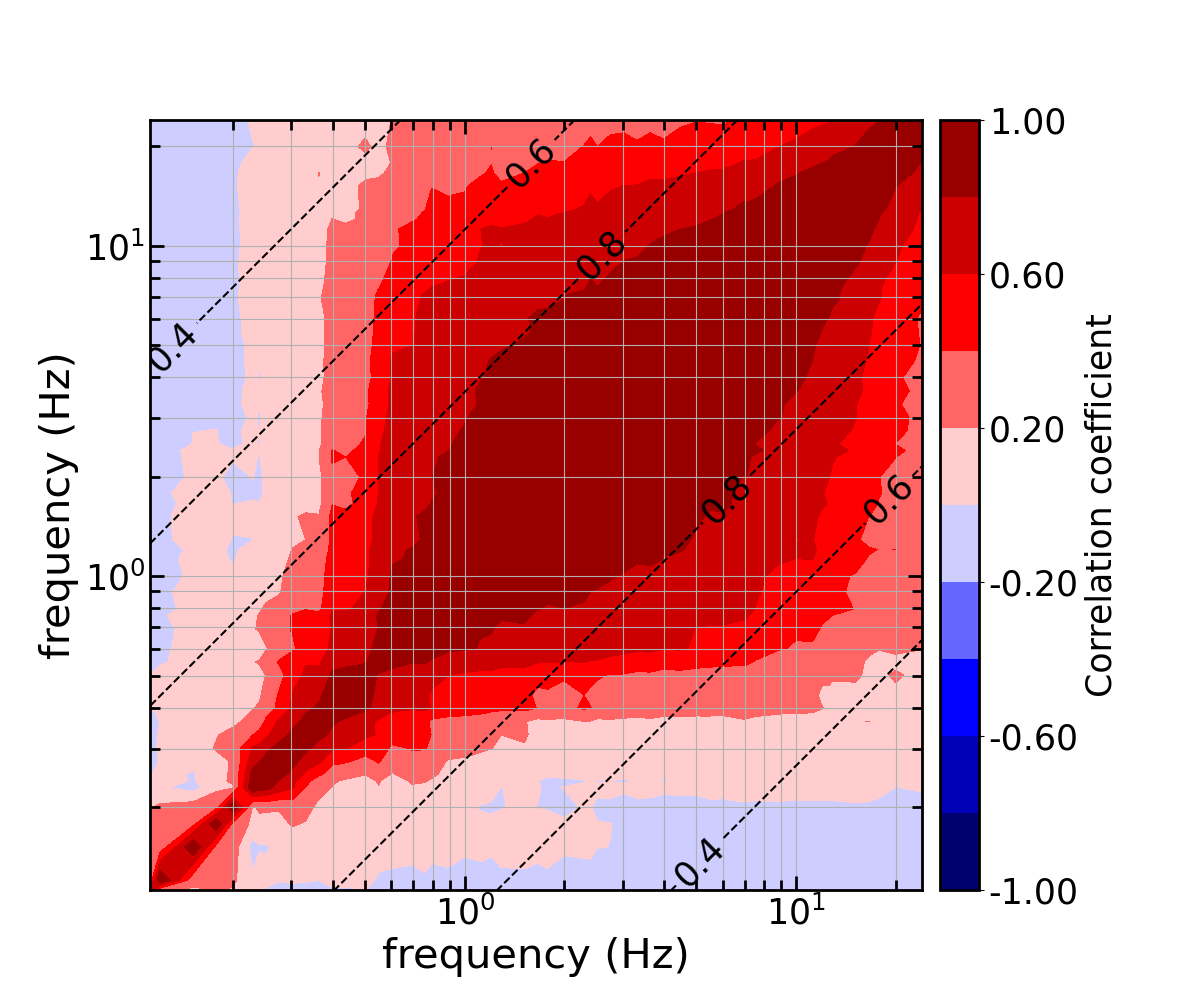}
    \end{subfigure}
    \caption{Inter-frequency correlation of non-ergodic terms; contour plot corresponds to the inter-frequency correlation of the data, dashed lines corresponds to the inter-frequency correlation model. 
    (a) inter-frequency correlation of $\delta c_{1,e}$,
    (b) inter-frequency correlation of $\delta c_{1a,s}$,
    (c) inter-frequency correlation of $\delta c_{1b,s}$, and
    (d) inter-frequency correlation of $c_{ca,p}$.}
    \label{fig:ifreq_corr}
\end{figure}

The inter-frequency correlation of the non-ergodic terms is presented in Figure \ref{fig:ifreq_corr} and the model parameters are summarized in Table \ref{tab:ifreq_corr_params}. 
Currently, the correlation of all non-ergodic terms is modeled as frequency independent: that is, the width of the EAS peaks and troughs does not depend their central frequency.
A frequency independent correlation would mean that the width of the correlation ridges in Figure \ref{fig:ifreq_corr} does not change along the diagonal, whereas a non-constant width would mean that the correlation changes with frequency. 
Out of all the non-ergodic terms, $\delta c_{1,e}$ has the widest confidence intervals because the number of unique earthquakes is smaller than both the number of unique stations or the number of anelastic attenuation cells. 
$\delta c_{1,e}$ has a relatively wide correlation, meaning that if $\delta c_{1,e}$ is positive at one frequency, it is highly likely that it will also be positive over a wide range of neighbouring frequencies.
The correlation of this term also exhibits some frequency dependence similar to the $\delta B_e$ frequency dependence found in \cite{Bayless2019}: the correlation is the widest at $f=0.5Hz$, it narrows at intermediate frequencies, and it widens again at frequencies larger than $8 Hz$. 
Both $\delta c_{1a,s}$ and $\delta c_{1b,s}$ have narrow, mostly frequency independent, inter-frequency correlations, which are similar to the correlation structure of $\delta {S2S}$ in \cite{Bayless2019}.  
The correlation of $c_{ca,p}$ shows the strongest frequency dependence; there is very little correlation at frequencies less than $1 Hz$, it gradually increases and reaches the widest point at $5 Hz$, and then, it narrows again. 
The narrow frequency correlation at low frequencies is expected as the anelastic attenuation is very weak for that frequency range; however, it is unclear why the inter-frequency correlation narrows at high frequencies.
It could be an artifact of poor sampling.
For now, it is modeled as frequency independent, but in future studies, this assumption would need to be reevaluated. 
As a point of comparison, in seismic numerical simulations, a deterministic velocity model would imply a perfect inter-frequency correlation on $c_{ca,p}$, which is more similar to the width of the correlation of data at $f=5Hz$

\begin{table}
	\caption {Interfequency model coefficients for non-ergodic terms}
	\centering
	\label{tab:ifreq_corr_params}
	\begin{tabular}{l c c c c} 
	    \hline\noalign{\smallskip}
	                        & A     & B     & C     & D     \\
    	\noalign{\smallskip}\hline\noalign{\smallskip}
	     $\delta c_{1,e}$   & 1.94  & 0.77  & 0.96  & 19.49 \\
	     $\delta c_{1a,s}$  & 1.30  & 0.92  & 1.36  & 30.85 \\
	     $\delta c_{1b,s}$  & 1.83  & 1.86  & 2.77  & 63.96 \\
	     $c_{ca,p}$         & 1.85  & 0.41  & 0.27  & 10.00 \\
	     \noalign{\smallskip}\hline
	 \end{tabular}
\end{table}

\subsection{Examples} \label{sec:examp}

\begin{figure}
    \centering
    \begin{subfigure}[t]{0.48\textwidth} 
        \caption{} \label{fig:eas_examp_ifreq}
        \includegraphics[width = .95\textwidth]{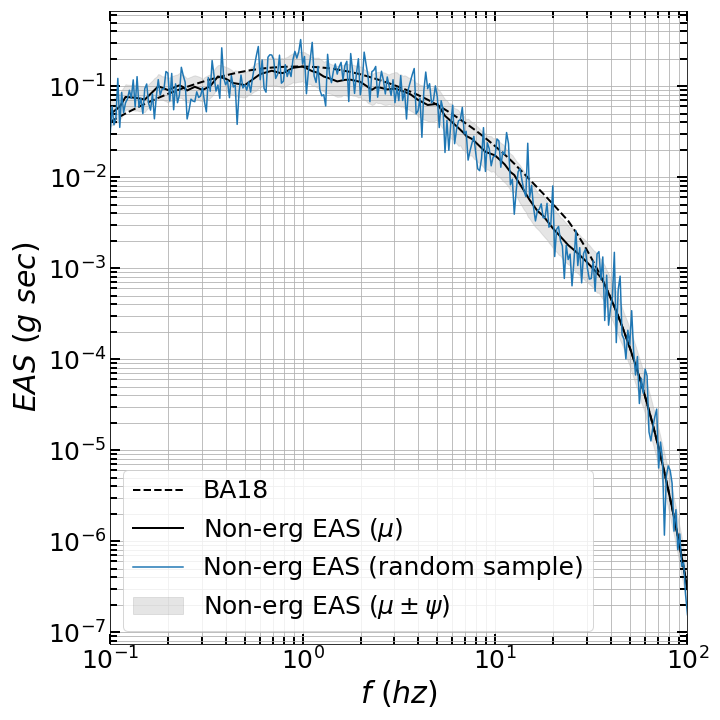}
    \end{subfigure}
    \begin{subfigure}[t]{0.48\textwidth}
        \caption{} \label{fig:eas_examp_cfreq}
        \includegraphics[width = .95\textwidth]{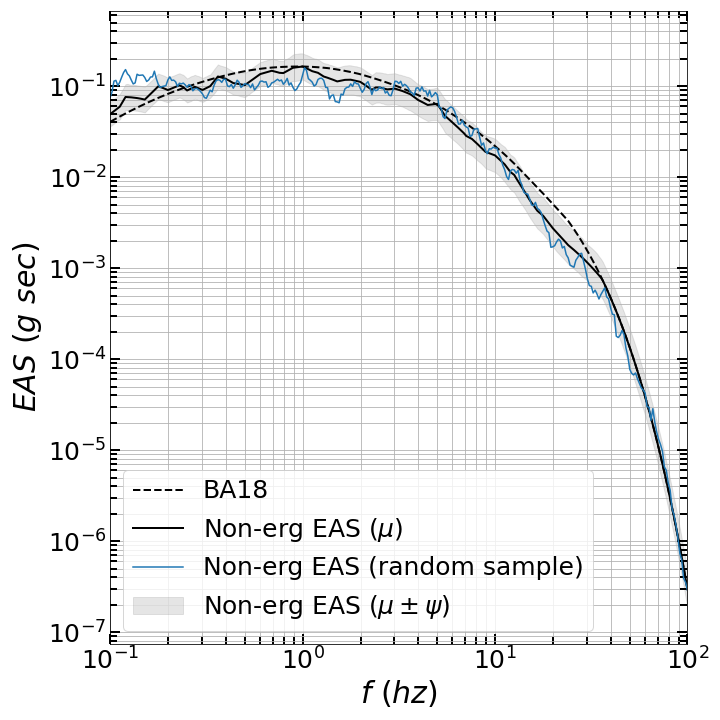}
    \end{subfigure}
    \caption{Effective amplitude spectra for a $M~7$ earthquake, $10~km$ away for a site located in Berkeley, CA. (a) without inter-frequency correlation, and (b) with inter-frequency correlation. }
    \label{fig:eas_examp}
\end{figure}

Figure \ref{fig:eas_examp} shows the effect of inter-frequency correlation in sampling the non-ergodic terms for an $M~7$ earthquake in Hayward fault $10~km$ from a site in Berkeley, CA.
Figure \ref{fig:eas_examp_ifreq} shows the median non-ergodic $EAS$, the $16^{th}$ to $84^{th}$ percentile range of epistemic uncertainty, and a representative $EAS$ sample with epistemic uncertainty for zero inter-frequency correlation. 
Figure \ref{fig:eas_examp_cfreq} shows the same information, but in this case, the ground motions were generated with the inter-frequency correlation model described previously. 
The median $EAS$ and range of epistemic uncertainty is the same in both cases; what is different are the representative EAS samples. 
The EAS sample with zero inter-frequency correlation has zero width in the peaks and the troughs, whereas EAS sample with inter-frequency correlation has peaks and troughs that span approximately a quarter of a decade.
It should be noted that these samples do not include aleatory variability, the inter-frequency correlation of the aleatory variability will influence the final width of the peaks and troughs. 

\begin{figure}
    \centering
    \begin{subfigure}[t]{0.48\textwidth} 
        \caption{} 
        \includegraphics[width = .95\textwidth]{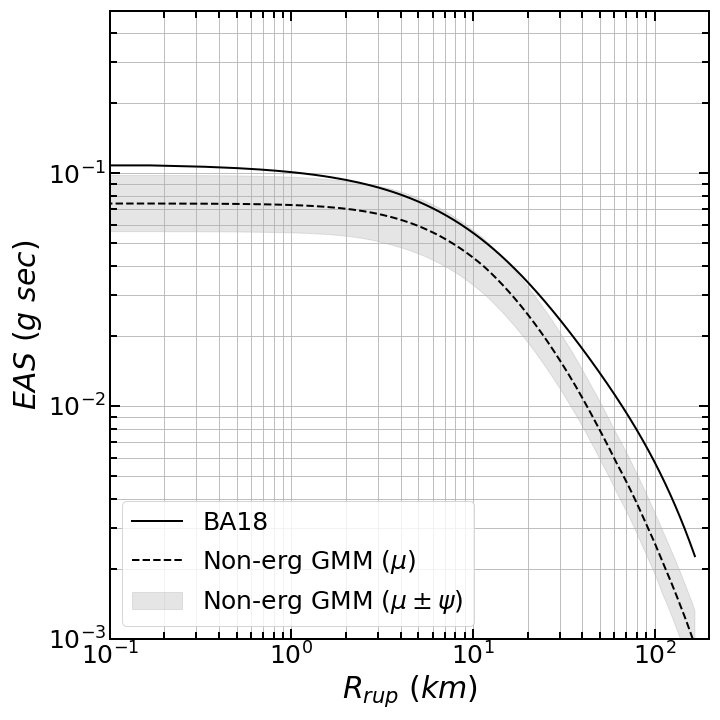}
    \end{subfigure}
    \begin{subfigure}[t]{0.48\textwidth}
        \caption{} 
        \includegraphics[width = .95\textwidth]{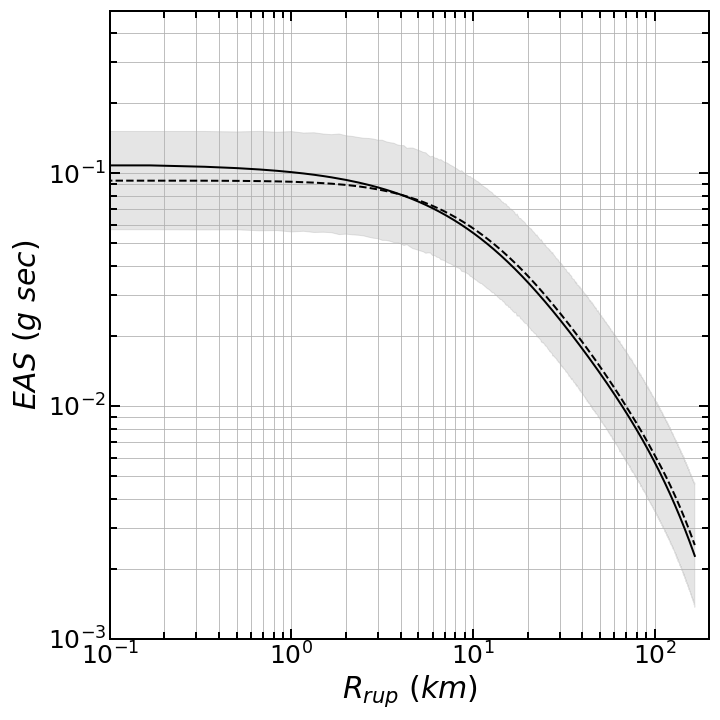}
    \end{subfigure}
    \caption{Distance scaling of $EAS(f=5Hz)$ for (a) a site in San Jose, CA and (b) a site in northeastern California.}
    \label{fig:dist_scl}
\end{figure}

The distance scaling of the model for $f=5Hz$ for a site in San Jose, CA ($SJ$) and a site in Northeastern California ($NE$) is presented in Figure \ref{fig:dist_scl}; the site in $SJ$ has a station which has recorded ground motions from past earthquakes to constrain $\delta c_{1b,s}$, whereas the site in $NE$ does not have one, so $\delta c_{1b,s}$ is unconstrained.
In both cases, the earthquakes are located north of sites.
North of $SJ$, $c_{ca,p}$ is less than average (Figure \ref{fig:cA_cell_mu}) which causes the non-ergodic GMM to have higher attenuation than BA18.
Due to the small number of paths in $NE$, $c_{ca,p}$ is very close to the mean value which is why the non-ergodic GMM and BA18 have similar distance scaling.
The epistemic uncertainty is less in $SJ$ as there are more earthquakes and stations to constrain the non-ergodic terms.

\section{Model Validation} \label{sec:valid}

\begin{figure}
    \centering
    \includegraphics[width = .45\textwidth]{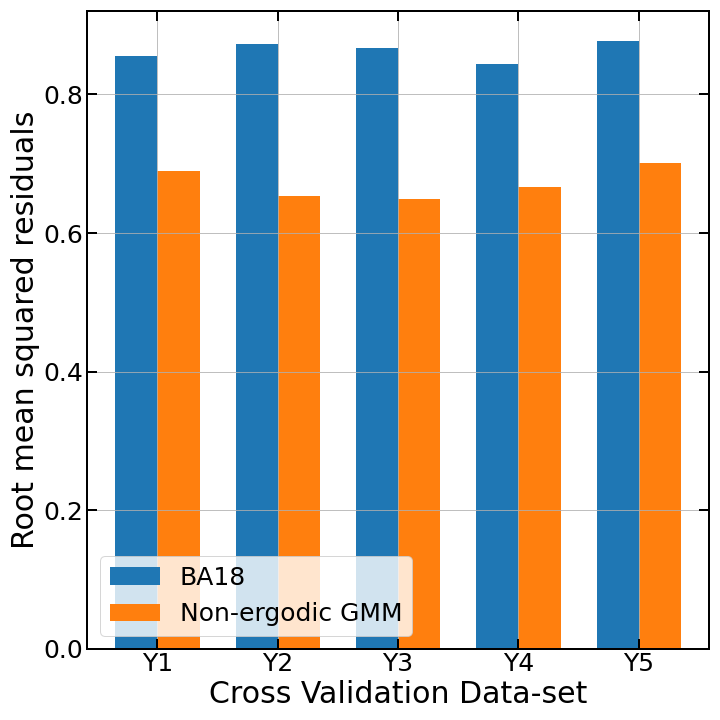}
    \caption{Root-mean-square error of 5-fold cross-validation test}
    \label{fig:CrossVal}
\end{figure}

The performance of the non-ergodic GMM was evaluated with a 5-fold cross validation test using the ground-motion data from $f=5hz$. 
In each of the $5$ iterations of the cross-validation test, the NGAWest2 CA dataset was randomly split into a training and test datasets, $80\%$ of the earthquakes composed training dataset and the remaining $20\%$ of earthquakes composed the test dataset.
The training set was used to estimate the coefficients of the non-ergodic model, and the test dataset was used to evaluate the accuracy of the predictions with the estimated coefficients. 
The NGAWest2 CA dataset was split based on the earthquakes so that the non-ergodic GMM is evaluated on events that were not used in the parameter estimation. 
Figure \ref{fig:CrossVal} shows the root-mean-square error ($rmse$) of the non-ergodic GMM and BA18 for all iterations.
The average $rmse$ of the non-ergodic GMM and BA18 is $0.66$ and $0.87$, respectively, which indicates that incorporating the non-ergodic terms improves the ground-motion prediction for events that were not part of the regression dataset.

\section{Conclusions and Discussion} \label{sec:conclusions}

A fully non-ergodic $EAS$ GMM is presented in this study. 
The non-ergodic source and station effects are captured by spatially varying coefficients; the non-ergodic path effects are captured with the cell-specific anelastic attenuation. 
A regional term that accounts for the differences in the ground motion of small earthquakes between northern and southern California is also added in the non-ergodic GMM; this term is applied to events less than $M~5$, and frequencies less than $5~Hz$.
The exact cause of these differences could not be identified, but it could be related to a potential bias in the magnitude estimation between the NCSN and SCSN networks.
Future studies should further investigate the cause of these differences. 

The proposed non-ergodic GMM has a  $30$ to $40\%$ smaller total aleatory standard deviation than BA18.
Furthermore, the cross-validation test shows that the non-ergodic GMM performs better than BA18 in predicting the ground motion for events that were not part of the regression dataset. 

The next step is to use this non-ergodic $EAS$ GMM with RVT to develop an equivalent non-ergodic $PSA$ GMM. 
The advantage of this approach is that it is easier to transfer the estimated non-ergodic terms, which are primarily based on small magnitude events, to the non-ergodic terms for the scenarios of interest, which typically are large magnitude events, using RVT than it is to estimate the magnitude dependence during the development of the non-ergodic $EAS$ GMM. 
For the scenarios of interest, the $PSA$ non-ergodic terms can be estimated by combining the $EAS$ predictions, for the same scenarios, with RVT. 
In this approach, the magnitude dependence of the non-ergodic $PSA$ terms is captured.

As larger data sets become available, future studies should consider the addition of a spatially varying term for geometrical spreading and test the frequency dependence of the inter-frequency correlation of the non-ergodic terms. 
A spatially-varying geometrical-spreading coefficient may be able to better capture the non-ergodic path effects at short distances; however, if such a coefficient is added, it should be constrained so that the GMM does not over-saturate at short distances. 
Currently, the inter-frequency correlation of the non-ergodic terms was assumed to be frequency independent, future studies should reevaluate if this assumption is valid.

Currently, the path for the cell-specific anelastic attenuation connects the site with closest point on the rupture. 
This path was chosen because it is the same path that is used in the $R_{rup}$ calculation; however, it has not been tested whether a path connecting the site and a different point on the rupture would be more appropriate for the cell-specific anelastic attenuation. 
A large number of broadband earthquake simulations that include $3D$ velocity structure effects up to high frequencies (e.g. 5 Hz) would be ideal for solving this problem.

\section{Acknowledgements}
This work was partially supported by the PG\&E Geosciences Department Long-Term Seismic Program.
The authors also thankful to the three anonymous reviewers for constructive comments that helped to improve the final article.

\section*{Declarations}
\subsection*{Funding}
This work was partially funded by the PG\&E Geosciences Department Long-Term Seismic Program.

\subsection*{Conflict of interest}
The authors declare that they have no conflict of interest.

\subsection*{Ethics approval}
Non applicable

\subsection*{Consent to participate}
Non applicable 

\subsection*{Consent for publication}
Non applicable 

\subsection*{Availability of data and material}

\subsection*{Code availability}
The are python scripts for the non-ergodic regressions are provided at:\\
{\fontfamily{qcr}\selectfont
https://github.com/glavrentiadis/NonErgodicGMM\_public
}

\bibliographystyle{chicago}
\bibliography{references_mendeley_GL.bib, references_other.bib} 

\end{document}